\documentclass[twocolumn]{aastex631}

\usepackage[T1]{fontenc}

\newcommand{\NII}{\makebox{[N{\sc\,II}]\,}}
\newcommand{\SII}{\makebox{[S{\sc\,II}]\,}}

\newcommand{\lyalpha}{Ly{$\alpha$}}
\newcommand{\hbeta}{H{$\beta$}}
\newcommand{\halpha}{H{$\alpha$}}
\def\CIV{C\,{\sc iv}}
\def\MgII{Mg\,{\sc ii}}
\def\FeII{Fe\,{\sc ii}}
\def\OIII{[O\,{\sc iii}]\,5007}
\def\OIIItwo{[O\,{\sc iii}]\,4959}
\def\HeII{[He\,{\sc ii}]\,4687}

\newcommand{\ps}{{\tt PrepSpec}}
\newcommand{\qsofit}{{\tt PyQSOFit}}
\newcommand{\brains}{{\tt BRAINS}}

\defcitealias{PancoastEtAl2014b}{P14}
\defcitealias{FriesEtAl2023}{F23}
\defcitealias{ShenEtAl2024}{S24}

\usepackage{float}

\usepackage{tabu}

\usepackage{changepage}
\usepackage{array}
\newcolumntype{?}{!{\vrule width 1.5pt}}

\usepackage{tabularx}
\newcommand\setrow[1]{\gdef\rowmac{#1}#1\ignorespaces}
\newcommand\clearrow{\global\let\rowmac\relax}

\shorttitle{Multi-Line Dynamical Modeling of a Highly Variable AGN}
\shortauthors{Stone et al.}
\graphicspath{{./}{figures/}}

\begin{document}

\title{The SDSS-V Black Hole Mapper Reverberation Mapping Project: Multi-Line Dynamical Modeling of a Highly Variable Active Galactic Nucleus with Decade-long Light Curves}

\author[0000-0002-8501-3518]{Zachary Stone}
\affiliation{Department of Astronomy, University of Illinois at Urbana-Champaign, Urbana, IL 61801, USA}
\email{stone28@illinois.edu}

\author[0000-0003-1659-7035]{Yue Shen}
\affiliation{Department of Astronomy, University of Illinois at Urbana-Champaign, Urbana, IL 61801, USA}
\affiliation{National Center for Supercomputing Applications, University of Illinois at Urbana-Champaign, Urbana, IL 61801, USA}

\author[0000-0002-6404-9562]{Scott F. Anderson}
\affiliation{Astronomy Department, University of Washington, Box 351580, Seattle, WA 98195, USA}

\author[0000-0002-8686-8737]{Franz Bauer}
\affiliation{Instituto de Astrof\'isica, Pontificia Universidad Cat\'olica de Chile, Av. Vicu\~na Mackenna 4860, Santiago, Chile}

\author[0000-0002-0167-2453]{W. N. Brandt}
\affiliation{Department of Astronomy \& Astrophysics, 525 Davey Lab, The Pennsylvania State University, University Park, PA 16802, USA}
\affiliation{Institute for Gravitation and the Cosmos, The Pennsylvania State University, University Park, PA 16802, USA}
\affiliation{Department of Physics, 104 Davey Lab, The Pennsylvania State University, University Park, PA 16802, USA}

\author[0000-0002-4469-2518]{Priyanka Chakraborty}
\affiliation{Center for Astrophysics|Harvard \& Smithsonian, Cambridge, MA}

\author[0000-0001-9776-9227]{Megan C. Davis}
\affiliation{Department of Physics, 196A Auditorium Road, Unit 3046, University of Connecticut, Storrs, CT 06269, USA}

\author[0000-0001-8032-2971]{Logan B. Fries}
\affil{Department of Physics, 196A Auditorium Road, Unit 3046, University of Connecticut, Storrs, CT 06269, USA}

\author[0000-0001-9920-6057]{Catherine~J.~Grier}
\affiliation{Department of Astronomy, University of Wisconsin-Madison, Madison, WI 53706, USA} 

\author[0000-0002-1763-5825]{P. B. Hall}
\affiliation{Department of Physics and Astronomy, York University, Toronto, ON M3J 1P3, Canada}

\author[0000-0002-6610-2048]{Anton M. Koekemoer}
\affiliation{Space Telescope Science Institute, 3700 San Martin Dr., Baltimore, MD 21218, USA}

\author[0000-0002-7843-7689]{Mary Loli Martínez-Aldama}
\affiliation{Astronomy Department, Universidad de Concepción, Casilla 160-C, Concepci\'on 4030000, Chile}

\author[0000-0002-4134-864X]{Knox Long} 
\affiliation{Space Telescope Science Institute, 3700 San Martin Dr., Baltimore, MD 21218, USA}

\author[0000-0002-6770-2627]{Sean Morrison}
\affiliation{Department of Astronomy, University of Illinois at Urbana-Champaign, Urbana, IL 61801, USA}

\author[0000-0001-5231-2645]{Claudio Ricci}
\affiliation{Instituto de Estudios Astrof\'isicos, Facultad de Ingenier\'ia y Ciencias, Universidad Diego Portales, Av. Ej\'ercito Libertador 441, Santiago, Chile}
\affiliation{Kavli Institute for Astronomy and Astrophysics, Peking University, Beijing 100871, People's Republic of China}

\author[0000-0001-7240-7449]{Donald P. Schneider}
\affiliation{Department of Astronomy and Astrophysics, The Pennsylvania State University, University Park, PA 16802, USA}
\affiliation{Institute for Gravitation and the Cosmos, The Pennsylvania State University, University Park, PA 16802, USA}

\author[0000-0001-8433-550X]{Matthew J. Temple}
\affil{Instituto de Estudios Astrof\'isicos, Facultad de Ingenier\'ia y Ciencias, Universidad Diego Portales, Av. Ej\'ercito Libertador 441, Santiago, Chile}

\author[0000-0002-1410-0470]{Jonathan R. Trump}
\affiliation{Department of Physics, 196A Auditorium Road, Unit 3046, University of Connecticut, Storrs, CT 06269, USA}



\begin{abstract}
We present dynamical modeling of the broad-line region (BLR) for the highly variable AGN SDSS J141041.25+531849.0 ($z = 0.359$) using photometric and spectroscopic monitoring data from the Sloan Digital Sky Survey Reverberation Mapping project and the SDSS-V Black Hole Mapper program, spanning from early 2013 to early 2023. We model the geometry and kinematics of the BLR in the \hbeta, \halpha, and \MgII\, emission lines for three different time periods to measure the potential change of structure within the BLR across time and line species. We consistently find a moderately edge-on $(i_{\rm full-state} = 53.29^{\circ} \,{}^{+7.29}_{-6.55})$ thick-disk $(\theta_{\rm opn, \; full-state} = 54.86^{\circ} \,{}^{+5.83}_{-4.74})$ geometry for all BLRs, with a joint estimate for the mass of the supermassive black hole (SMBH) for each of three time periods, yielding $\log_{10}(M_{\rm BH} / M_{\odot}) = 7.66^{+0.12}_{-0.13}$ when using the full dataset. The inferred individual virial factor $f$ $\sim 1$ is significantly smaller than the average factor for a local sample of dynamically modeled AGNs. There is strong evidence for non-virial motion, with over $80\%$ of clouds on inflowing/outflowing orbits. We analyze the change in model parameters across emission lines, finding the radii of BLRs for the emission lines are consistent with the following relative sizes $R_{\rm H\beta} \lesssim R_{\rm MgII } \lesssim R_{\rm H\alpha}$. Comparing results across time, we find $R_{\rm low-state} \lesssim R_{\rm high-state}$, with the change in BLR size for \hbeta\, being more significant than for the other two lines. The data also reveal complex, time-evolving, and potentially transient dynamics of the BLR gas over decade-long timescales, encouraging for future dynamical modeling of fine-scale BLR kinematics. 
\end{abstract}



\section{Introduction} \label{sec:intro}

It is well-known that nearly all massive galaxies harbor a supermassive black hole (SMBH) at their center \citep{MagorrianEtAl1998, KormendyHo2013}. Properties and effects of these SMBHs are most commonly observed through actively accreting SMBHs known as active galactic nuclei (AGNs). AGNs have been observed for nearly a century \citep{Seyfert1943, Schmidt1963}, and display a number of common properties in their time-series and spectroscopic data, including a characteristic stochastic variability on all timescales and observed wavelengths, broad emission lines, narrow emission lines, and continuum emission \citep{Netzer2013}. To explain observations of these different parts of the spectrum, a general picture of the inner emitting regions of AGNs has been put forward \citep[e.g.,][]{Antonucci1993, UrryPadovani1995, Netzer2015b}. The geometry and kinematics of these substructures imprint themselves on their emission, and reflect the properties of the SMBH, their effects on the SMBH's surroundings, and their relation to the rest of its host galaxy. AGN temporal variability, and its correlation between different photometric bands, has been modeled well over year-long timescales with both statistically-informed \citep[e.g.,][]{KellyEtAl2009a, MacLeodEtAl2010a, KellyEtAl2014, SuberlakEtAl2021, StoneEtAl2022b} and physically-informed \citep[e.g.,][]{DexterAgol2010a, SunEtAl2020a, NeustadtKochanek2022d, StoneShen2022a} models. Such studies have demonstrated the connection between variability and physics within the AGN, including the timescale of optical variability correlating positively with the mass of the SMBH across a wide range of masses \citep{BurkeEtAl2021}, and with the rest-frame observed wavelength \citep[e.g.,][]{MacLeodEtAl2010a}.

However, there are many parts of this picture that remain uncertain, including the different geometries of the inner emitting substructures: the accretion disk, broad-line region (BLR), narrow-line region, dusty torus, etc. Statistical prescriptions employed to describe AGN optical variability have been shown to produce biased results \citep[e.g.,][]{Kozlowski2017} and break down on days- to weeks-long timescales when using decades-long light curves \citep{StoneEtAl2022b}. Work on finer scales within the AGN can reveal the nature of certain substructures and probe properties of the AGN and SMBH. In particular the BLR, a region of gas clouds orbiting the SMBH photoionized by a central source to produce broad emission lines, has been frequently used to infer the mass of the SMBH $M_{\rm BH}$ through a method known as reverberation mapping \citep[RM;][]{BlandfordMcKee1982a, Peterson1993, Peterson2001}.

RM exploits the intrinsic variability of AGNs, correlating the emission from the central source, assumed to arise from the accretion disk, to emission from the BLR which echoes the continuum \citep{Peterson2001}. Recombination times for the measured broad lines are relatively short, meaning path-dependent effects are dominant, so the emission line can be treated as a delayed and smoothed version of the photoionizing continuum radiation \citep{Peterson2001}. Assuming a virialized BLR, RM relates $M_{\rm BH}$ to a measured time delay $\tau$:

\begin{equation}
    M_{\rm BH} = f\frac{c \tau v^2}{G}
\end{equation}

\noindent where $c$ is the speed of light, $G$ is the gravitational constant, $v$ is a measured velocity of the BLR gas, and $f$ is a normalization factor, accounting for the geometry and kinematics of the BLR \citep{Peterson2001}. 

$\tau$ can be obtained through time-series correlation methods and $v$ measured via the width of a given emission line. $f$ can be calibrated in a variety of ways, including with the empirical $M_{\rm BH}-\sigma_*$ relation observed in the local Universe \citep{OnkenEtAl2004a, GrahamEtAl2011, ParkEtAl2012, WooEtAl2013a, HoKim2014}, but only for the population average rather than for individual systems. Combining RM with the empirical $R-L$ relation \citep{KaspiEtAl2000, BentzEtAl2009, BentzEtAl2013} allows the calculation of $M_{\rm BH}$ with only a single spectrum \citep[e.g.,][]{WandelEtAl1999, Vestergaard2002, McLureJarvis2002, VestergaardPeterson2006, Shen2013, ShenEtAl2016, ShenEtAl2024}. Large-scale RM survey programs, such as the Sloan Digital Sky Survey Reverberation Mapping \citep[SDSS-RM;][]{ShenEtAl2015} project and the OzDES Reverberation Mapping program \citep[e.g.,][]{KingEtAl2015, HoormannEtAl2019} obtain spectra for a large number of broad-line AGNs at multiple epochs to obtain lags and $M_{\rm BH}$ over a wide range of redshifts and AGN properties. 

\begin{figure*}
    \centering
    \includegraphics[width=\textwidth]{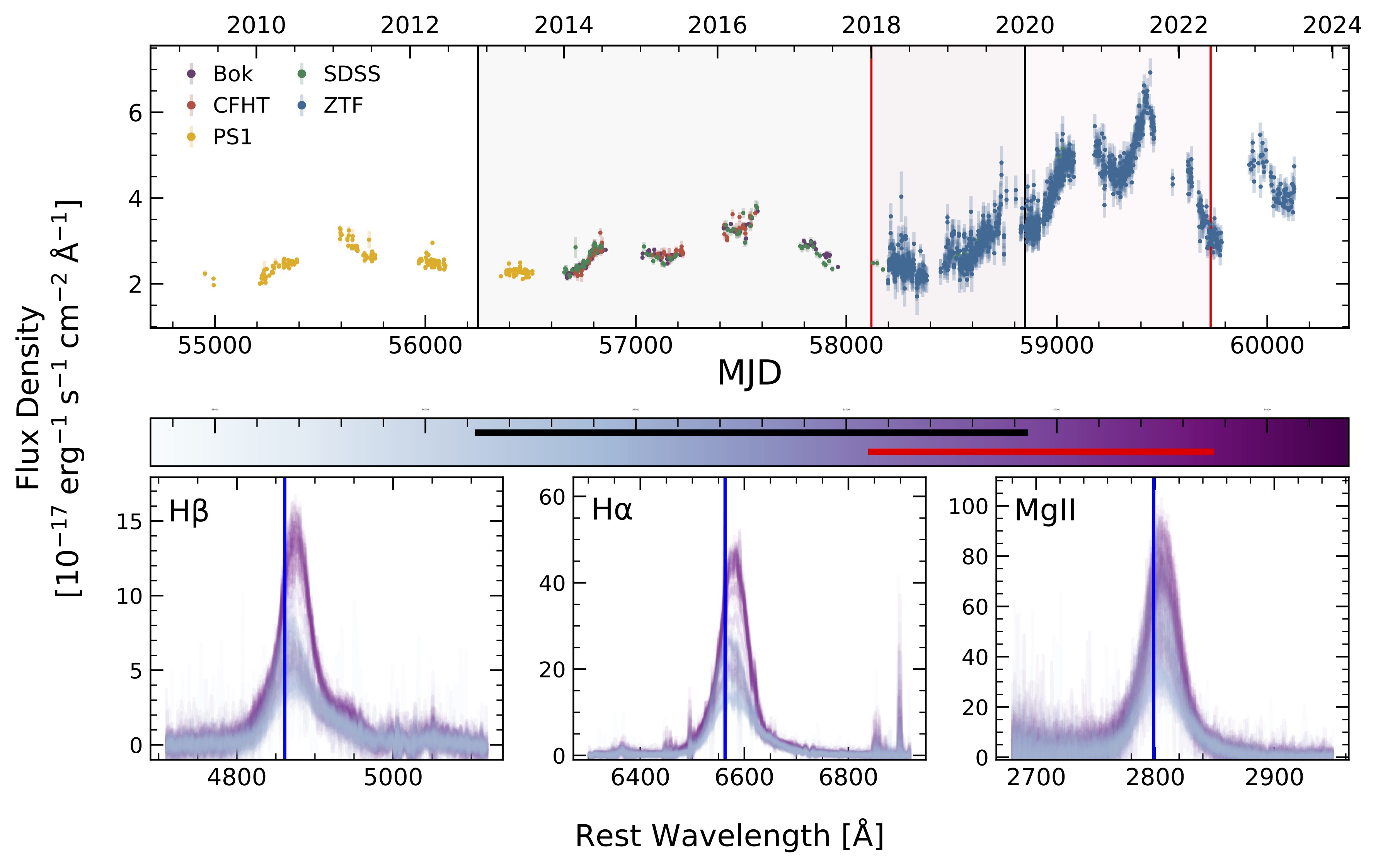}
    \caption{\emph{Top:} The continuum light curve for RM160, color-coded by data set. The time period for the low(high) state is shown shaded in black(red). \emph{Bottom:} The \qsofit-processed, calibrated line profiles for all observations of the lines, color-coded by the observed MJD. The colorbar has the same date range as the continuum panel, and the low(high) state range is shown with the bold black(red) line. The systemic redshift of RM160 is based on the peak of \OIII. We leave overlap in the two states to improve statistical constraints. }
    \label{fig:data}
\end{figure*}

There are many aspects of general $M_{\rm BH}$ estimation that could be significantly improved: the uncertainty in a single-epoch measurement can be $\sim 0.3-0.4$~dex due to the uncertainty carried in $f$ and the $R-L$ relation \citep[e.g.,][]{ShenEtAl2024}. To improve estimates of $f$, and investigate its dependency on conditions within the BLR and properties of the AGN, independent measurements of $M_{\rm BH}$ are needed. Traditional RM makes many assumptions about the BLR, including virial motions and temporal stability. Furthermore, most RM work only measures a mean lag of the BLR, although this can be improved by employing velocity-resolved RM to map out kinematics across the observed line profile. Finally, traditional RM assumes the width of a given emission line is constant, whereas it is known that the profile of the broad line may change over time due to changes in physical conditions within the BLR along or out of our line-of-sight (LOS) \citep[e.g.,][]{KoristaGoad2004, BarthEtAl2015, WangEtAl2020}.

To accurately map the geometry and kinematics of the BLR requires derivation of the transfer function $\Psi(v,\tau)$ that governs the response of the BLR to the photoionizing continuum \citep{HorneEtAl2004a}, as a function of LOS velocity and time delay. There are multiple methods to infer BLR geometry and kinematics given the transfer function, including velocity-resolved reverberation mapping \citep[e.g.,][]{DenneyEtAl2009, BentzEtAl2009, GrierEtAl2013}, the maximum entropy method \citep[MEMEcho;][]{WelshHorne1991, HorneEtAl2004a}, linear inversion \citep{SkielboeEtAl2015}, and dynamical modeling \citep[][hereafter \citetalias{PancoastEtAl2014b}]{BottorffEtAl1997, PancoastEtAl2014b}. In particular, dynamical modeling is an appealing method to probe the transfer function, as it utilizes a physically-motivated dynamical model to obtain estimates for all physical parameters involved, including $M_{\rm BH}$ and BLR inclination. By mapping the kinematics of the gas within the BLR, dynamical modeling can test  the virial assumption in RM and produce $\Psi(v,\tau)$ as a byproduct. Thus, dynamical modeling can jointly infer both the size of the BLR and $M_{\rm BH}$, without requiring certain important assumptions in deriving the BH mass (e.g., viriality and inclination) as in traditional RM work.


Over the past decade, nearly 30 AGNs have been analyzed with dynamical modeling using RM data, displaying a general trend in BLR geometries: a thick disk, inclined relatively face-on toward the observer \citep{PancoastEtAl2014, GrierEtAl2017c, PancoastEtAl2018, LiEtAl2018, WilliamsEtAl2018a, WilliamsEtAl2020a, WilliamsEtAl2021, WangEtAl2020a, VillafanaEtAl2022a, BentzEtAl2021a, BentzEtAl2022, BentzEtAl2023}. Dynamical modeling results reveal a variety of BLR kinematics, including inflow, outflow, virial motion, or a combination of them. In most instances, the results obtained from dynamical modeling match those obtained from traditional RM within their uncertainties, even with the measurement of $M_{\rm BH}$ from dynamical modeling being independent of the assumptions present in traditional RM.

More objects have been analyzed through their velocity-resolved lag profiles, including samples of super-Eddington AGNs \citep{DuEtAl2016}, and those with asymmetric broad emission line profiles \citep{BrothertonEtAl2020, BaoEtAl2022, UEtAl2022, ZastrockyEtAl2024}. Generally, AGNs containing broad emission lines with red asymmetry produce lag profiles representative of inflow or virial motion with some influence of inflow, and those with blue asymmetry produce lag profiles representative of outflow or virial motion with some influence of outflow. Although, there are a number of objects that counter this notion, with either contradictory lag profiles or complex profiles requiring more elaborate BLR models to explain \citep[e.g.,][]{DuEtAl2018}. Super-Eddington AGN lag profiles also display a range of BLR kinematics \citep{DuEtAl2016}. The inferred kinematics from velocity-resolved RM and dynamical modeling agree in a number of AGNs \citep[e.g.,][]{VillafanaEtAl2022a}.


\begin{table*}
    \centering
    \caption{\brains\, Parameter Descriptions}

    \begin{adjustwidth}{-2.4cm}{}
    \begin{tabular}{|p{.08\textwidth}|p{.35\textwidth}|p{.075\textwidth}?p{.08\textwidth}|p{.3\textwidth}|p{.075\textwidth}|}
        \hline
        Parameter & Description & Bounds & Parameter & Description & Bounds \\
        \hline
        $R_{BLR}$ & Mean radius of the BLR & & $\sigma_{\rho,rad}$ & Std. dev. of the $v_r$ distribution for non-elliptical orbits & \\[-2pt]
        $\beta$ & Shape parameter for the radial cloud distribution &  & $\sigma_{\theta,rad}$ & Std. dev. of the $v_\phi$ distribution for non-elliptical orbits & \\
        $F$ & Inner radius of the BLR &  & $\theta_e$ & Location of the center of the cloud velocity distribution in the $v_r - v_\phi$ plane & [0, 90]$^{\circ}$ \\
        $i$ & \tablenotemark{\footnotesize \rm \textbf{I} }Inclination angle of the system & [0, 90]$^{\circ}$ & $\sigma_{turb}$ & Std. dev. of the macroturbulent velocity distribution & \\[-2pt]
        $\theta_{opn}$ & \tablenotemark{\footnotesize \rm \textbf{II} }Disk opening angle & [0, 90]$^{\circ}$ & $\Delta V_{line}$ & Instrumental line broadening & \\[-9pt]
        $\kappa$ & \tablenotemark{\footnotesize \rm \textbf{III} }Near-side/Far-side preference parameter & [-0.5, 0.5] & $\sigma_{sys,line}$ & Systematic line error & \\[-10pt]
        $\gamma$ & \tablenotemark{\footnotesize \rm \textbf{IV} }Disk outer face clustering parameter& & $\sigma_{sys,con}$ & Systematic continuum error & \\[-10pt]
        $\xi$ & \tablenotemark{\footnotesize \rm \textbf{V} }Midplane transparency parameter & [0, 1] & $\sigma_d$ & Damped Random Walk (DRW) long-term variability amplitude & \\
        $M_{BH}$ & Mass of the SMBH & & $\tau_d$ & DRW timescale & \\[-11pt]
        $f_{ellip}$ & Fraction of clouds on elliptical orbits & [0, 1] & B & Linear trend slope & \\[-10pt]
        $f_{flow}$ & \tablenotemark{\footnotesize \rm \textbf{VI} }Inflow/Outflow parameter & [0, 1] & $A$ & Response amplitude & \\[-10pt]
        $\sigma_{\rho,circ}$ & Std. dev. of the $v_r$ distribution for elliptical orbits & & $A_g$ & Response power-law index & \\
        $\sigma_{\theta,circ}$ & Std. dev. of the $v_\phi$ distribution for elliptical orbits & & & & \\
        \hline
    \end{tabular}
    \end{adjustwidth}
    \vspace{3pt}
    \begin{adjustwidth}{0cm}{}
        \hspace{6pt} $^{\rm \textbf{I}}$ $i = 0^{\circ}$ is face-on, $i=90^{\circ}$ is edge-on \hspace{105pt} $^{\rm \textbf{IV}}$ Larger $\gamma$ suggests more clustering \\
        \hspace{1.5pt} $^{\rm \textbf{II}}$ $\theta_{opn} = 0^{\circ}$ is a disk, $\theta_{opn} = 90^{\circ}$ is a sphere \hspace{83.5pt} $^{\rm \textbf{V}}$ $\xi = 0$ is opaque, $\xi = 1$ is transparent \\
        $^{\rm \textbf{III}}$ $\kappa < 0$ is near-side, $\kappa > 0$ is far-side \hspace{108pt} $^{\rm \textbf{VI}}$ $f_{flow} < 0.5$ is inflow, $f_{flow} > 0.5$ is outflow
    \end{adjustwidth}
    \label{tab:param_descriptions}
\end{table*}


However, almost all dynamical modeling has been performed using \hbeta\, data, except for a handful of AGNs for which dynamical modeling has been performed for UV broad lines such as \CIV\, \citep{WilliamsEtAl2020a, WilliamsEtAl2021}. In general, it has been shown that different emission lines respond to the continuum differently \citep[e.g.,][]{ClavelEtAl1991, KoristaGoad2004, SunEtAl2015, HomayouniEtAl2020, WangEtAl2020}. Emission lines can arise from a variety of physical processes, such as the Balmer lines resulting from recombination, while a significant amount of radiation from \MgII\, stems from collisional excitation. Thus, their BLR distances and structures may be different. Photoionization calculations demonstrate that the BLR is stratified, with low-ionization lines (LILs; e.g., \hbeta, \MgII, and \halpha) having longer lags than high-ionization lines (HILs; \, e.g., \CIV\, and \lyalpha) \citep{GoadEtAl1993, OBrienEtAl1995, KoristaGoad2000}. Indeed, early RM studies confirmed that HIL lags are shorter than LIL lags \citep{ClavelEtAl1991, GoadEtAl1999a, PetersonWandel2000, KollatschnyEtAl2001, Kollatschny2003}. HILs have been shown to be more variable than LILs \citep{GoadEtAl1999a, GoadEtAl1999, SunEtAl2015}, and extend over a smaller radius \citep{GuoEtAl2020}. \citet{WilliamsEtAl2020a} performed dynamical modeling on multiple lines, analyzing \hbeta, \lyalpha, and \CIV\, for NGC 5548. They find evidence for stratification, with the \CIV\ and \lyalpha\, BLRs being more compact and less radially extended, and the \hbeta\, BLR located farther out and spread over a larger range of radii. In addition, they report similar geometries for all line-emitting regions across the BLR. \citet{BentzEtAl2021a} published similar results dynamically modeling high-ionization \HeII\, and low-ionization \hbeta\, for NGC 3783.

\MgII\, has been a line of particular interest, as it is the most readily accessible broad line for quasars at $1<z<2$ with optical spectroscopy, the epoch that includes the bulk of black hole growth \citep[e.g.,][]{BrandtAlexander2015}. It is well-known that \MgII\, is less responsive to the continuum than the Balmer lines, as shown by photoionization \citep[e.g.,][]{GoadEtAl1993, OBrienEtAl1995, KoristaGoad2000, GuoEtAl2020} and variability \citep[e.g.,][]{SunEtAl2015, YangEtAl2020a, WangEtAl2020} studies. RM surveys have demonstrated that \MgII\, lags are longer than or similar to those for \hbeta\, \citep{HomayouniEtAl2020, YuEtAl2021, YuEtAl2023, ShenEtAl2024}, although these lags are more difficult to obtain due to the weaker response and longer temporal baselines needed. Many studies find similarities between the two lines: \citet{YuEtAl2023} and \citet{HomayouniEtAl2020} find the \MgII\, $R-L$ relation slope matches that of the \hbeta\, $R-L$ relation, although the \MgII\, relation has more scatter. However, based on {\tt CLOUDY} \citep{ChatzikosEtAl2023} modeling, \citet{GuoEtAl2020} conclude that there is no clear $R-L$ relation for \MgII. While in many instances the observational tracers of the BLRs for the Balmer lines and \MgII\, are similar, different physical processes may lead to different BLR geometries and kinematics for both types of lines.

SDSS J141041.25+531849.0 (hereafter RM160) at $z=0.359$ is a broad-line quasar within the SDSS-RM sample, showing continuum variations of more than one magnitude, and large response following the continuum in its broad emission lines. RM160 has a measured bolometric luminosity $\log_{10}(L_{\rm bol} / {\rm erg s^{-1}}) = 44.50$, Eddington ratio $\lambda_{\rm Edd} = 0.015$ \citep{ShenEtAl2019}, and SMBH mass via \hbeta\, $\log_{10}(M_{\rm BH} / M_{\odot}) = 8.11$ \citep{ShenEtAl2024}.  The emission line variability of RM160 has recently been analyzed in \citet[][hereafter \citetalias{FriesEtAl2023}]{FriesEtAl2023}, and shown to exhibit stratification, line breathing \citep{KoristaGoad2004} in \halpha, \hbeta, and \MgII, and possible azimuthal asymmetries. RM160 has high signal-to-noise ratio (S/N), multi-epoch spectra spanning nearly 10 years in the observed frame, containing prominent, varying broad emission lines, including \halpha, \hbeta, and \MgII. In addition, RM160 has been shown to experience two different continuum states (a low and high state), during which Fries et al. (2024, in prep) suggest a transition due to change in BLR structure. 

To obtain an in-depth physically-motivated look at the complex BLR of RM160, in this work we analyze the reverberation data of RM160 via dynamical modeling of \hbeta, \halpha, and \MgII. In Section~\ref{sec:data} we present the RM data for RM160 and describe the dynamical modeling method. In Section~\ref{sec:results} we interpret the model fits for all lines and states, compare to prior work, and investigate the change in the virial factor $f$ with model parameters. Finally, we conclude in Section~\ref{sec:conclusion}. In Appendix~\ref{sec:caveats} we describe caveats inherent in the dynamical model. In Appendix~\ref{sec:method_comp}, we compare dynamical modeling results between two different spectral reduction methods. Throughout this work, all mentions of the time lag $\tau$ will be in the rest frame of the AGN.

\section{Data}\label{sec:data}

We utilize spectroscopic monitoring data from the SDSS-RM project \citep{ShenEtAl2015} and the current fifth generation of SDSS \citep[SDSS-V;][]{SDSS-V}, in combination with photometric monitoring data from various sources \citep[compiled by][]{ShenEtAl2024}. SDSS-RM was a multi-year multi-object spectroscopic RM program, obtaining 11 years of photometric monitoring data (2010-2020) and 7 years of spectroscopic monitoring data, across both SDSS-III \citep{SDSS-III} and SDSS-IV \citep{SDSS-IV}, using some early photometric data from the PanSTARRS-1 \citep[PS1;][]{PS1} survey. SDSS-RM obtained data for 849 quasars using the BOSS multi-fiber spectrographs \citep{BOSS-Spec} on the Apache Point Observatory 2.5m SDSS telescope \citep{SDSS-APO}, obtaining 90 epochs of spectroscopy during 2014-2020. Within SDSS-V, we utilize data from the Black Hole Mapper Reverberation Mapping (BHM-RM) program, which has continued to obtain RM data for 380 AGNs in the original SDSS-RM sample, and hundreds more in additional fields from BHM-RM. The BHM-RM program has been obtaining data since 2021, providing a maximum baseline of 10 years for spectroscopic data and 15 years for photometric data, particularly appealing for objects in the 7 deg$^2$ SDSS-RM field, such as RM160.

\subsection{Photometry}

Within SDSS-RM, RM160 was observed photometrically for 876 nightly epochs over 11 years with the Steward Observatory Bok 2.3m telescope on Kitt Peak and the 3.6m Canada-France-Hawaii Telescope (CFHT) on Maunakea across different optical bands. We obtain additional photometry from public Zwicky Transient Facility \citep[ZTF;][]{ZTF} data, spanning four more years up to late 2023. We utilize the merged continuum light curve from SDSS-RM and the \emph{gri} band light curves from recent ZTF data, combining them using {\tt PyCALI} \citep{LiEtAl2014}, scaling the output light curve to the SDSS \emph{r}-band, and finally converting to $\rm f_\lambda$. We then coadd intra-night epochs from the output merged light curve weighted by their inverse variance. The resulting continuum light curve is shown in Fig.~\ref{fig:data}.

As described in \citetalias{FriesEtAl2023}, there is a significant increase in the continuum emission and variability starting in late 2020 - early 2021, increasing in flux by over 150\%. We separate the data and modeling into two separate periods based on the variability seen in the continuum: the low state (01/2013 - 01/2020), and the high state (01/2018 - 06/2022). We intentionally keep some overlap between these two periods to improve the statistical constraints for both states. This differs slightly from the definition of states presented in Fries et al. (2024, in prep), who define the low state as 2013-2019 and the high state as 2021-2023, separating the states by two years. We label the full state as the union of both states (01/2013 - 06/2022). The average BLR lag of RM160 is much shorter than a single season, so we limit the time period of analysis from 2013 - 2022 to better match the spectroscopic baseline (\citetalias{FriesEtAl2023}). This selection produces 1924 nightly-coadded photometric epochs over a period of 14 years.

\subsection{Spectroscopy}\label{subsec:spectroscopy}

We obtain spectroscopy from both the SDSS-RM project and SDSS-V survey. For SDSS-III and SDSS-IV data, we utilize the same procedure to obtain data outlined in Fries et al. (2024, in prep), but only using data after 11/2013. For more recent data of RM160, we search the ``master'' version of the compiled summary file for SDSS-V (``spAll*.fits'') as of January 2024, using the fiber coordinates of a given observation, a search range of 0\farcs3, and a time range spanning our full state definition. In total, we have 138 epochs over a spectroscopic baseline of eight years. This spectroscopic dataset excludes the first few sparse epochs used for the emission line analysis in \citetalias{FriesEtAl2023} in early 2013, and adds a few in mid-2022.

Before any spectral fitting, we calibrate the flux of the spectra with \ps\, (see \citealp{ShenEtAl2015, ShenEtAl2016} for a description). We input the raw (whole) spectra into \ps, performing full fits for each epoch. \ps\, outputs the decomposed elements of the spectrum for each epoch, including a continuum, narrow/broad lines, as well as calibration factors to account for the change in conditions across observations. \ps\, calibrates the raw spectra such that the shift and scale of the \OIII\, line is equal for each epoch. We utilize these flux factors as calibration factors for our processed line profiles. To verify our results, we also use a process similar to \citetalias{FriesEtAl2023} to calibrate our spectra and find consistent results with \ps, although the spectra calibrated with \ps\, show less scatter in \OIII\, parameters.

After flux calibration with \ps, we utilize \qsofit\, to fit our spectra to retrieve decomposed broad-line spectra. \qsofit\, \citep{GuoEtAl2018} is a $\chi^2$-based spectral fitting software that decomposes an input AGN spectrum into different components, including host-galaxy flux, \FeII\, flux, a power-law/polynomial continuum, narrow emission lines, and broad emission lines. \qsofit\, also employs Monte Carlo sampling to produce uncertainties in fit parameters.

\qsofit\, fits for the host-galaxy contribution using principal component analysis. We first correct for Galactic reddening in the input spectra (see \citealp{ShenEtAl2019}), then fit the host-galaxy contribution over the entire spectral range for each epoch separately, allowing the model to include a \FeII\, contribution and power-law continuum but not any emission lines. We choose to use 20 components for the AGN spectrum and five components for the host-galaxy spectrum, using eigenspectra from \citet{YipEtAl2004, YipEtAl2004a}, as this selection produced the most accurate model fits. The median of all-epoch host-galaxy contributions was selected as the ``true'' host-galaxy flux and was subtracted from all epochs before further analysis. 

The underlying local continuum was subtracted from each of the three considered lines (\halpha, \hbeta, \MgII) separately. We perform a joint fit with both a power-law continuum and \FeII\, emission template \citep{BorosonGreen1992, VestergaardWilkes2001}, using the line-fitting bounds described in Section 3 of \citet{ShenEtAl2011a}. This combined \FeII+power-law component was fit using an iterative procedure:
\begin{enumerate}
    \item We perform a single fit per epoch to obtain initial estimates of the velocity shift, full-width at half maximum (FWHM), and amplitude of the \FeII\, template
    \item We then refit all epochs where one of these three \FeII\, parameters are $ > 5\sigma$ from the median of this parameter across all epochs, fixing whichever parameter to the median
    \item If, after this fit, a particular epoch's \FeII\, parameters are still $ > 5\sigma$ from the median, we fix both the FWHM and the velocity shift
    \item If an epoch's \FeII\, parameters remain $> 5\sigma$ from the median, we fix the \FeII\, parameters to the median over all epochs
\end{enumerate}
We then subtract the \FeII+power-law continuum from each epoch for each line, leaving a calibrated, continuum-free line profile.

\begin{figure}
    \centering
    \includegraphics[width=.5\textwidth]{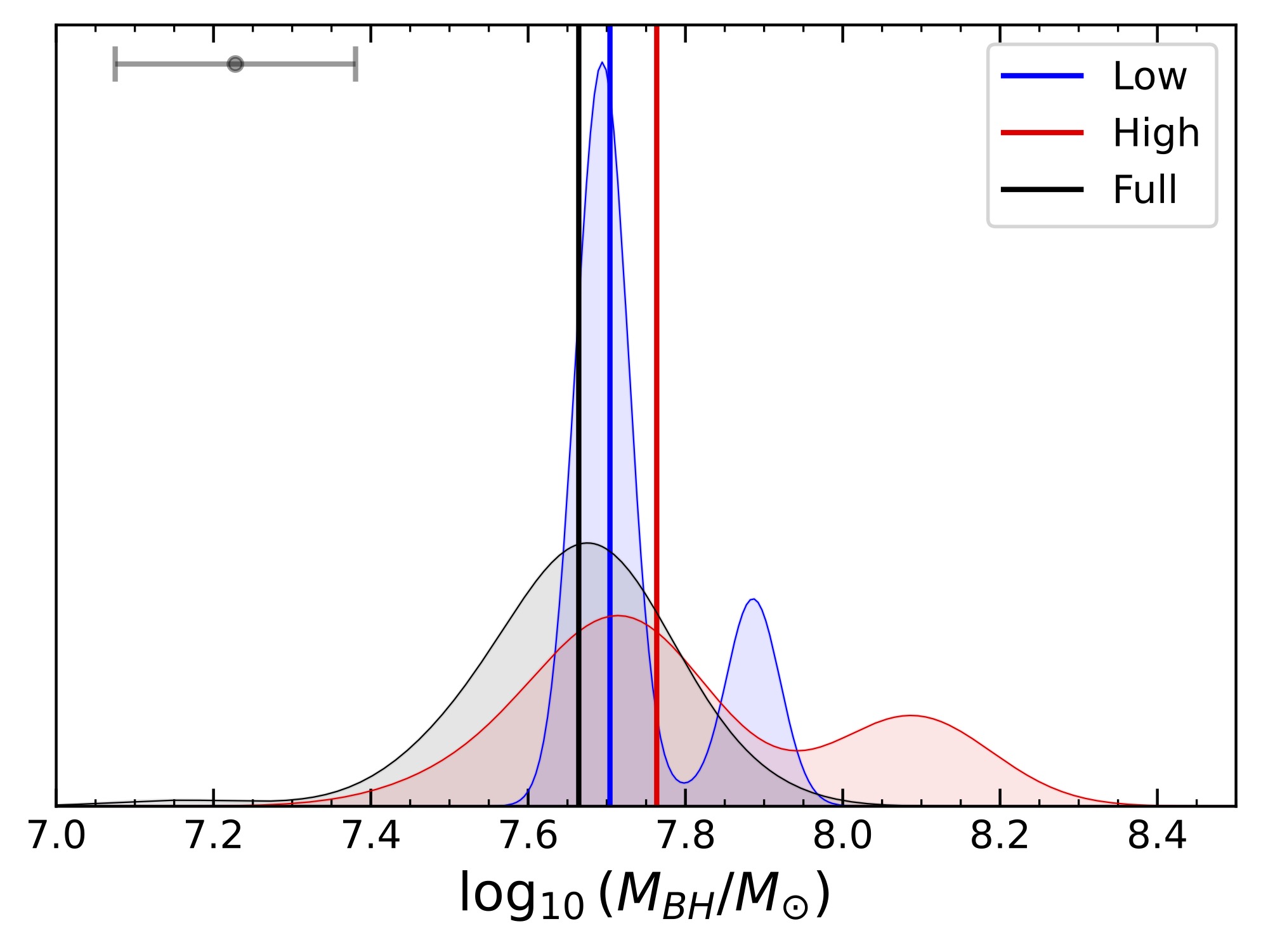}
    \caption{The joint posterior (i.e., combining the posteriors for each of the 3 lines - $\rm P_{H\alpha} \cdot P_{H\beta} \cdot P_{MgII}$) for $M_{\rm BH}$ for each state. Each posterior has been produced using a Gaussian kernel density estimate (KDE). Vertical lines represent the median value for the corresponding posterior. The gray data point in the top left represents the average error in the median estimate across all states.}
    \label{fig:mbh_posterior}
\end{figure}

We perform joint broad/narrow-line fits of each epoch and each line separately. We first fit the \hbeta\, profiles across all epochs, using three Gaussians to fit the broad line, one Gaussian to fit the narrow \hbeta\, line, and one broad and one narrow line to fit the broad and narrow components of \OIII\, and \OIIItwo. We constrain the amplitude of \OIIItwo\, to be one third of the \OIII\, amplitude \citep[e.g.,][]{StoreyZeippen2000}. We restrict the width and shift of the broad line [O\,{\sc iii}] components, as well as \HeII, to avoid overlap with broad \hbeta, which produces more accurate fits. We anchor the narrow line fits in the \halpha\, fitting process to have the same width and velocity shift as \OIII, as the properties of \OIII\, have been shown to be fairly constant over periods of years \citep{PetersonEtAl1982}, as well as for RM160 over the monitoring period \citepalias{FriesEtAl2023}. We fit the \halpha\, broad line with three Gaussians, the \halpha\, narrow line with one Gaussian, and the \NII and \SII doublets with two Gaussians each. If the fits are poor to the \halpha\, profile, we relax the restriction of matching to the \OIII\, profile. For \MgII, we fit a broad component with two Gaussians, and a narrow component with one Gaussian not tied to \OIII. This approach produces profiles for the broad and narrow components within each line. To acquire the broad line profile, we subtract the fitted narrow line profile from the continuum-subtracted line profile.

\subsection{Dynamical Modeling}\label{subsec:dynamical_modeling}

\begin{figure*}
    \centering
    \includegraphics[width=\textwidth]{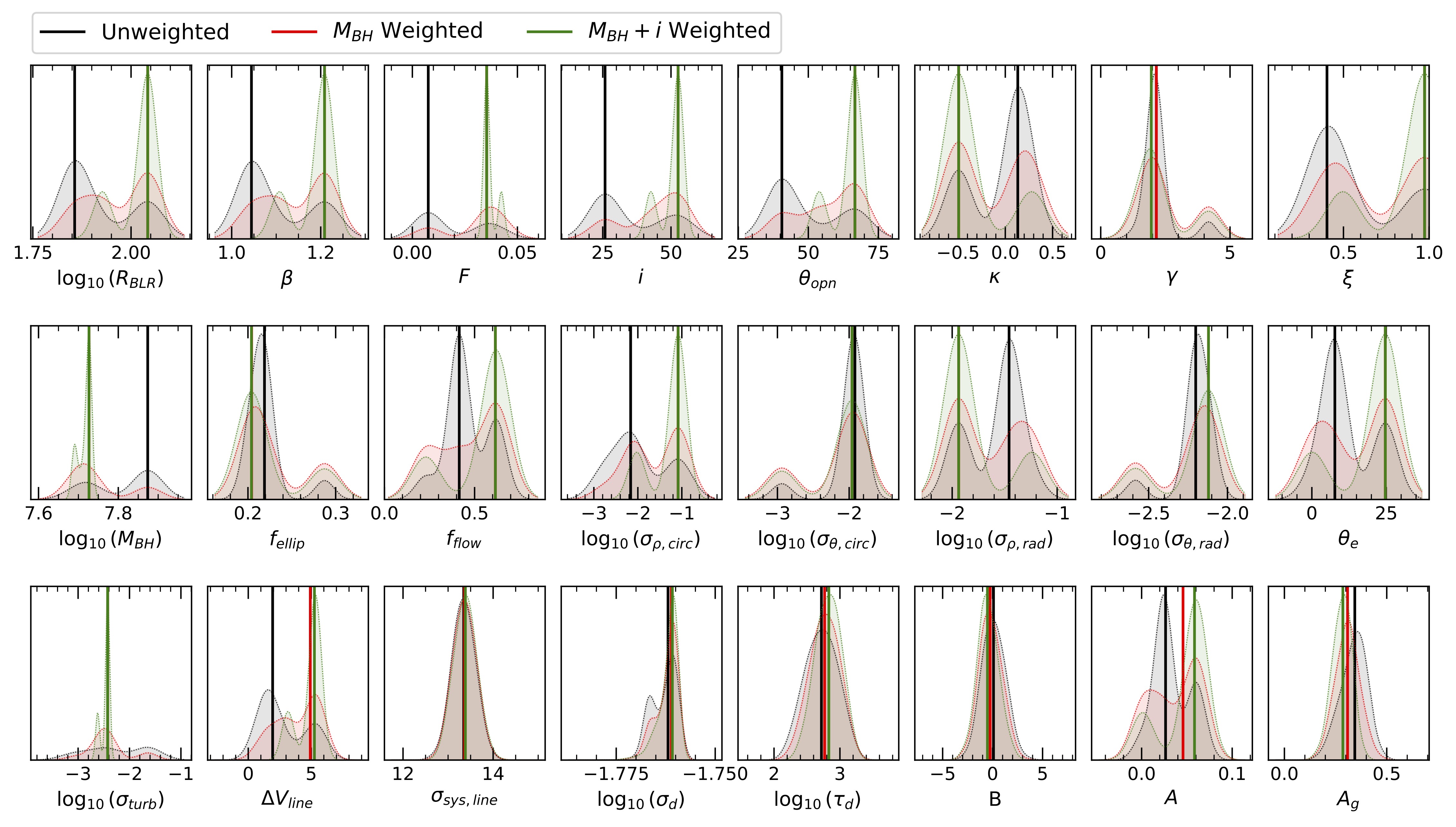}
    \caption{The output posterior distributions for all \brains\, parameters, using different weighting schemes, for \halpha\, in the low state. Posteriors have been produced with a Gaussian KDE. Vertical lines in each of the panels represent the weighted median values from the posterior samples for a given line. We opt to use the $M_{\rm BH}$-weighted posterior, shown in red, as it enforces the same $M_{\rm BH}$ across lines and sufficiently constrains the posterior distribution.}
    \label{fig:posterior_comp}
\end{figure*}

To perform dynamical modeling, we use \brains\, \citep{LiEtAl2013, LiEtAl2018}, assigning the geometric and kinematic model described in \citetalias{PancoastEtAl2014b}. For a more in-depth description of the model and Bayesian framework used for the fitting of spectral and continuum data see \citet{LiEtAl2018} and \citetalias{PancoastEtAl2014b};  we will provide a brief description here. For figures depicting the model, see Fig.~1 of \citet{LiEtAl2013} and Fig.~2 of \citetalias{PancoastEtAl2014b}. For a discussion of caveats within the model, see Appendix~\ref{sec:caveats}.


\brains\, uses a Bayesian framework to maximize the likelihood between the model-produced and the input (observed) continuum and spectra. Each set of \brains\, model parameters governs the radial and azimuthal distribution of clouds and their velocities. Each cloud is assigned a velocity governed by parameters in the model. The clouds are allowed to be on elliptical or outflowing/inflowing orbits, both of which are treated differently via different model parameters. It should be noted -- non-elliptical (i.e., inflowing/outflowing) orbits can be either bound or unbound. Clouds on non-elliptical orbits will have a velocity distribution centered around the escape velocity. Consequently, a portion of the non-elliptical orbits will still be bound, although highly elliptical. Thus, whenever outflowing/inflowing clouds are mentioned, a portion of those treated as inflowing/outflowing are still bound on Keplerian orbits. These clouds absorb continuum photoionizing radiation from the central source close to the SMBH, whose mass is governed by a model parameter, and immediately radiate it toward the observer. Thus, for each set of clouds, a transfer function $\Psi$ is produced, which can be used to create a set of model spectra given a model continuum. Additionally, \brains\, has a number of parameters to account for asymmetry within the geometry, including the transparency of the midplane and emissive weights. Finally, \brains\ accounts for both a trend in the continuum and line light curves, as well as non-linear response in the line. These parameters, their descriptions, and their bounds are listed in Table~\ref{tab:param_descriptions}, but also in \citet{LiEtAl2018}. \brains\, then uses the diffusive nested sampling \citep{BrewerEtAl2009a} software {\tt CDNest} \citep{Li2018} to sample the likelihood space.

\begin{figure*}
    \centering
    \includegraphics[width=\textwidth]{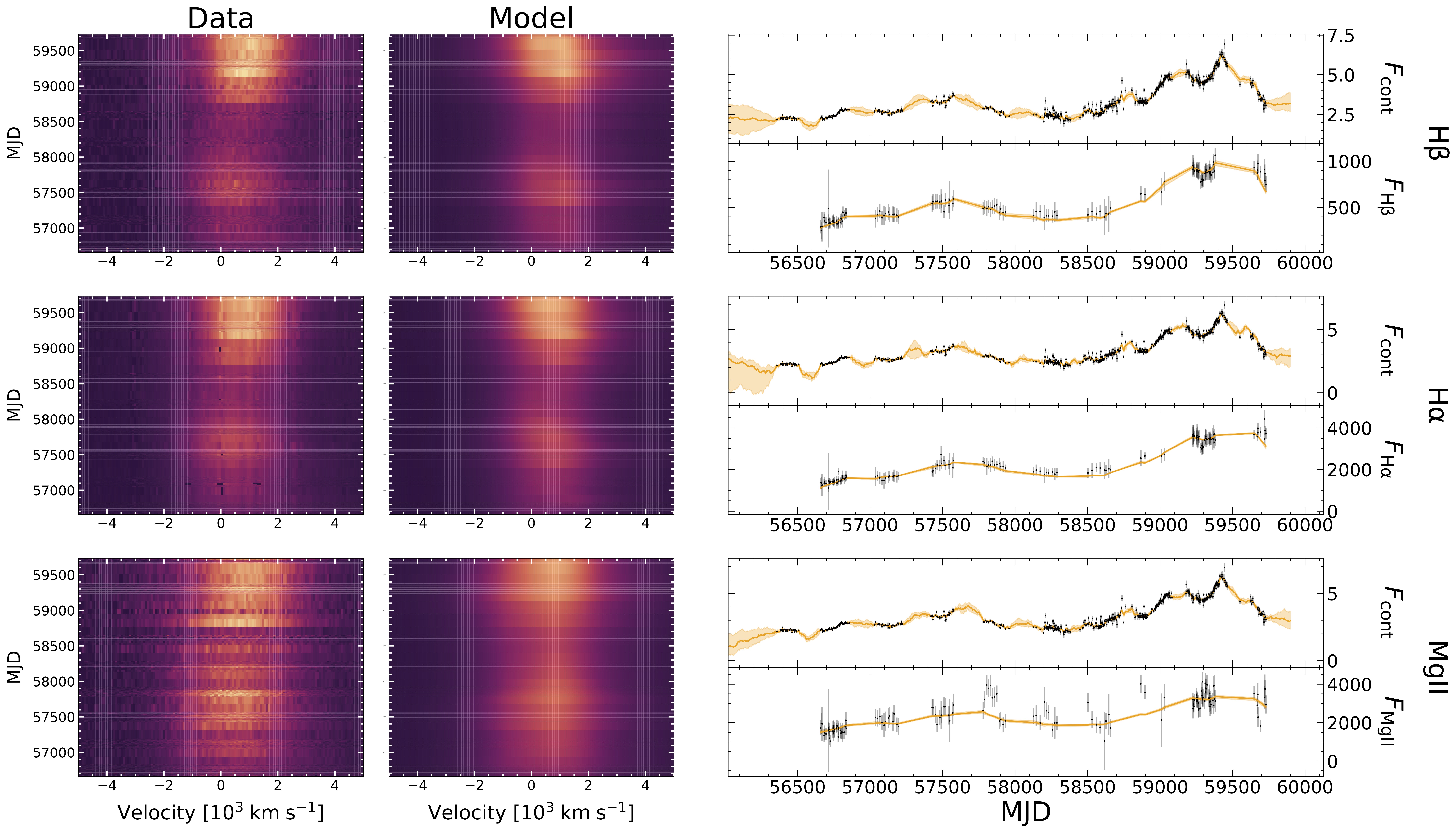}
    \caption{Comparison of {\tt BRAINS} output for the full state for all lines. \emph{Left:} Input and model-reconstructed multi-epoch spectra. The model spectra are generated using the maximum (weighted) probability parameters. \emph{Right:} Fits to the continuum and line light curve data from the model. The weighted median light curve fit is shown as an orange line, while the weighted $16^{th} - 84^{th}$ percentile range of all posterior sample light curve fits are shaded in light orange. The error bars for the line light curves are adjusted using the weighted median $\sigma_{sys,line}$ and the temperature used $(\sqrt{T})$.}
    \label{fig:brains_res_tot1}
\end{figure*}

As input, \brains\, requires both a continuum light curve and multi-epoch spectra for each run. In order to determine the range of wavelengths to consider, we utilize the \qsofit\, model fits to the broad lines for each line profile. We define a tolerance such that the wavelength range for a given epoch is set when the broad line model reaches this value on either side. We choose a conservative value of $5 \times 10^{-19}\; \rm erg \; s^{-1} \; cm^{-2}$ \AA$^{-1}$ for this tolerance to provide the most information possible to \brains. Since \brains\, requires a common set of wavelength bins for each epoch, we bin the line-profile data. Approximately 1\% of the wavelength bins contain no spectral data points; for these bins we interpolate the line profiles using B-splines implemented in {\tt scipy} \citep{scipy}. We assign these interpolated points an uncertainty using the standard deviation of the surrounding 10 data points. 

\brains\, has a number of ``hyper-parameters'' that must be set before
starting a given run. For each run, we produce 50000 clouds per core, with each cloud being assigned five velocities. We opt not to include a systematic uncertainty added to the continuum to reduce degeneracy, and set a maximum radius for clouds in the BLR of 500 light-days. This maximum lies well outside the dust sublimation radius for RM160 of $\approx 128$ light-days, assuming the ultraviolet bolometric correction factor in \citet{Netzer2019} and Eqn.~5 in \citet{Barvainis1987}. We continue each \brains\, run until 28000 samples have been produced. 

\begin{figure*}
    \begin{interactive}{js}{interactive_figure/Fig5.html}
        \includegraphics[width=\textwidth]{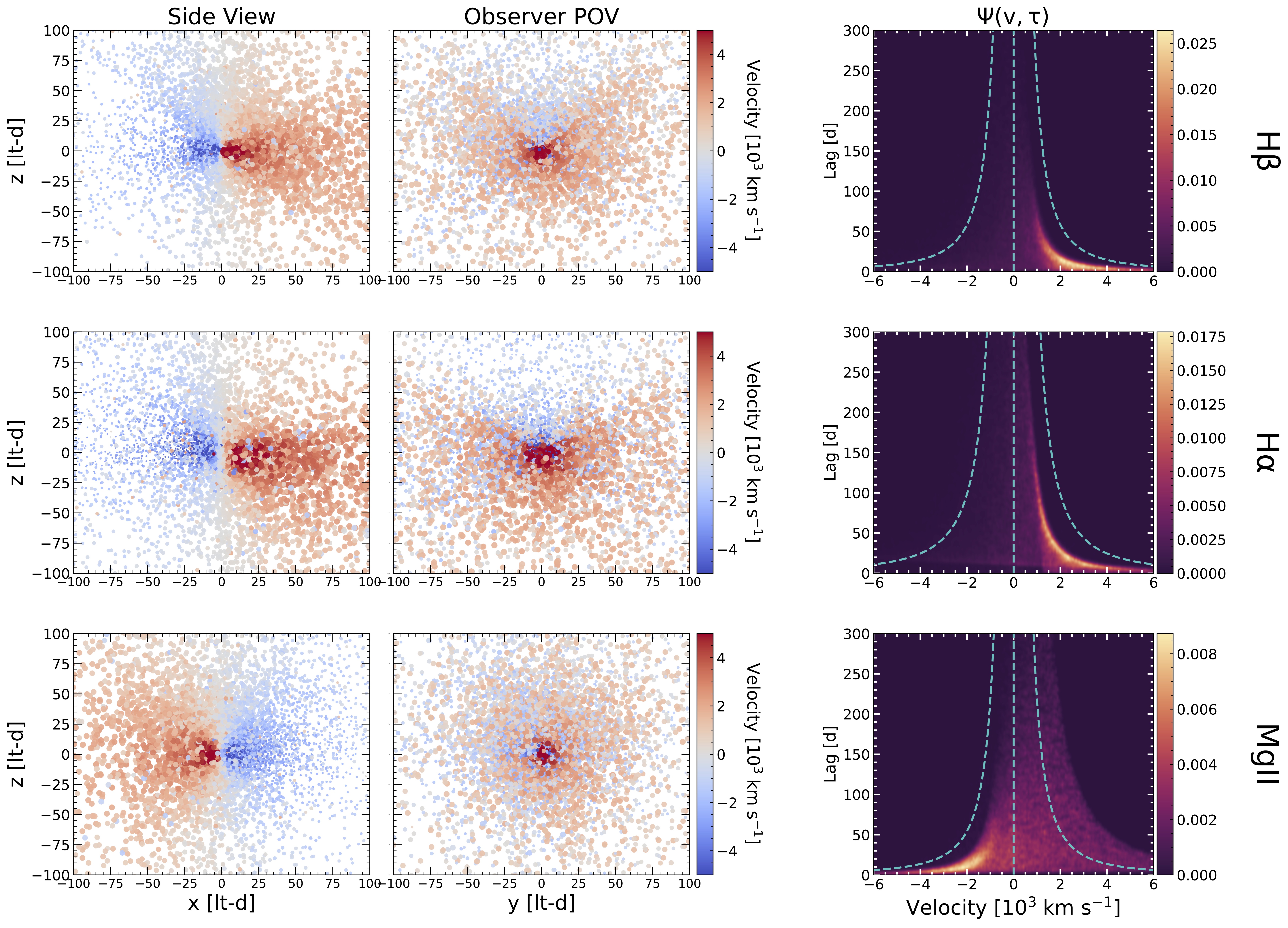}
    \end{interactive}
    \caption{Comparison of {\tt BRAINS} output for the full state for all lines. \emph{Left/Top:} Cloud configurations generated by the best-fit parameters for each run. Two points of view (POVs) are shown for each cloud configuration, along and perpendicular to the observer's LOS. The clouds are colored by their LOS velocity to the observer. Cloud size is assigned by its emissive weight, determined by the model parameter $\kappa$. \emph{Right/Bottom:} Transfer functions $\Psi(v,\tau)$ generated using the best-fit model parameters for each line. The cyan dashed line represents the virial envelope, using the best-fit $M_{\rm BH}$, as well as $v=0$. Many models produce transfer functions with an asymmetric distribution, concentrated near large positive (i.e., red) velocity shifts, due to the large fraction of inflowing BLR clouds. A {\tt Plotly} interactive version of this figure is available in the online Journal.}
    \label{fig:brains_res_tot2}
\end{figure*}

After the run has finished, obtaining posterior samples is performed as a post-processing step. For more information on posterior sampling within diffusive nested sampling methods, see \citet{BrewerEtAl2009a} and \citet{Li2018}. In general, when obtaining these samples, a temperature $T$ is chosen to soften the likelihood, effectively increasing the measurement uncertainties by a factor $\sqrt{T}$. We first obtain the $T$ value for each run that grants at least 1000 posterior samples. We then increase the temperature until reaching a maximum number of samples, if the value of $T$ is not large (in general $T < 200$). The output posteriors can be sensitive to the temperature used, so we avoid using large temperatures (i.e., uncertainties). We use a sufficiently high temperature to yield a sufficient number of posterior samples, as the smaller the number of posterior samples produced, the sparser posterior samples are distributed within the model parameter-space.

We perform dynamical modeling of \hbeta, \halpha, and \MgII\, \qsofit-reduced data in the low state, high state, and the combination of both states which we will label the ``full state''. As a check, we also perform dynamical modeling using the \ps-reduced spectra; the results between the two are similar (see Appendix~\ref{sec:method_comp}). 

\section{Results}\label{sec:results}


Many of the model parameters display continuous posterior distributions for each of the runs. However, in some instances, there are outliers present in the posteriors with high likelihoods, causing the best-fit parameters, and thus the best-fit model configuration, to not be representative of the entire posterior. There are a number of combinations of parameters that fit the variability seen in the continuum and broad emission lines. To choose a set of best-fit model parameters, we follow the posterior weighting procedure described in \citet{WilliamsEtAl2020a} and \citet{BentzEtAl2021a}. Briefly, this approach weights regions of higher density in the joint $M_{\rm BH}$ posterior (Fig.~\ref{fig:mbh_posterior}) across all lines for a given state more, to ensure that $M_{\rm BH}$ is consistent for a given state. For each state, we produce a posterior for $M_{\rm BH}$ for each line, then combine them to produce a joint posterior. Employing importance sampling \citep[see e.g., Appendix B of][]{LewisBridle2002}, the weighting for a given posterior sample is the ratio of the individual $M_{\rm BH}$ posterior for a given line and state to the joint $M_{\rm BH}$ posterior for that same state. This weighting better constrains the parameter posteriors, and makes them more consistent across lines. For completeness, we weight by the joint $M_{\rm BH} + i$ posterior; this approach constrains the posteriors to too small of a region within the model parameter-space. A comparison of posteriors using these different weighting schemes is shown in Fig.~\ref{fig:posterior_comp}. We opt to use the $M_{\rm BH}$-weighted posteriors for this analysis unless otherwise stated. 

\begin{figure*}
    \centering
    \includegraphics[width=\textwidth]{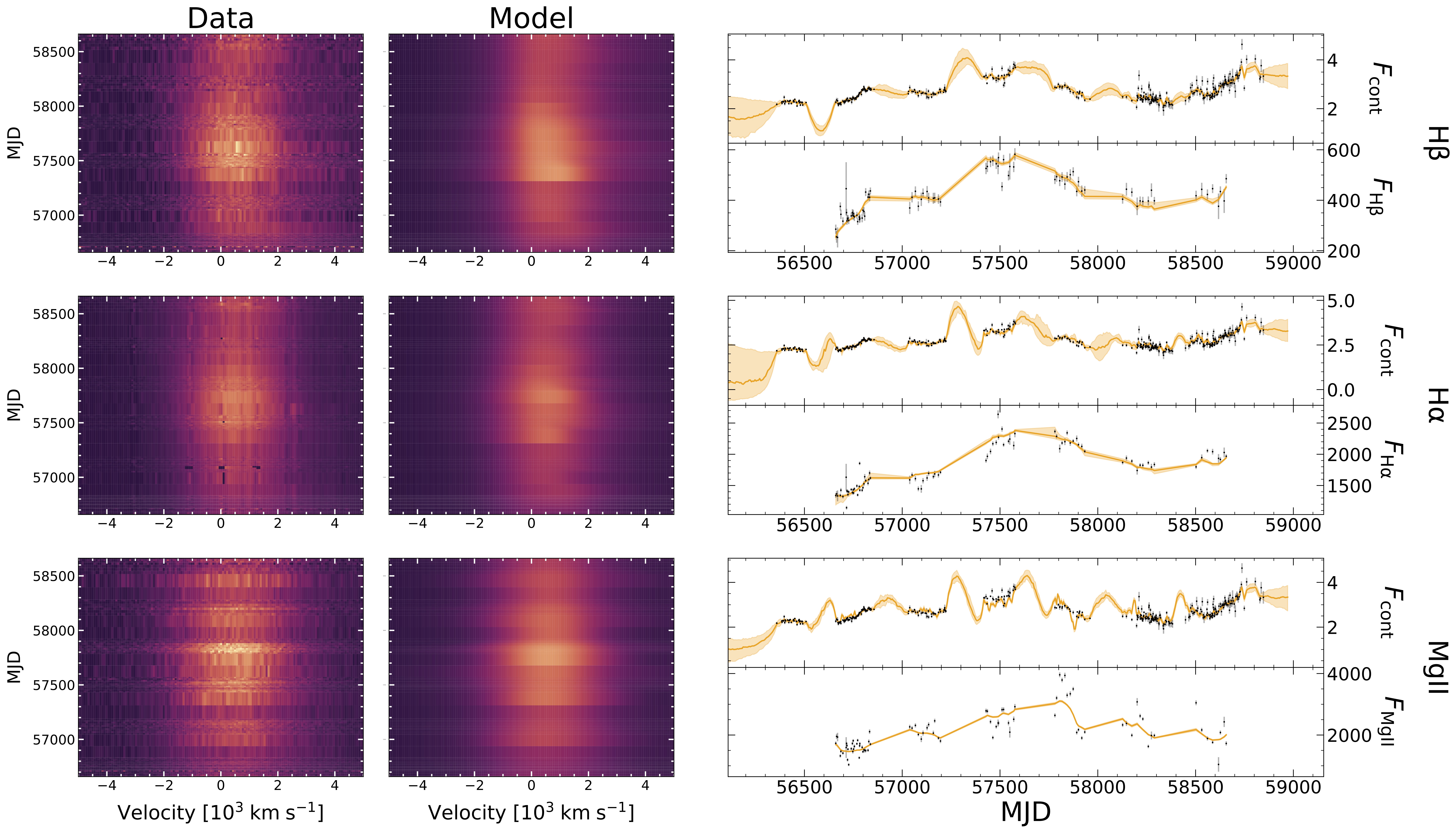}
    \includegraphics[width=\textwidth]{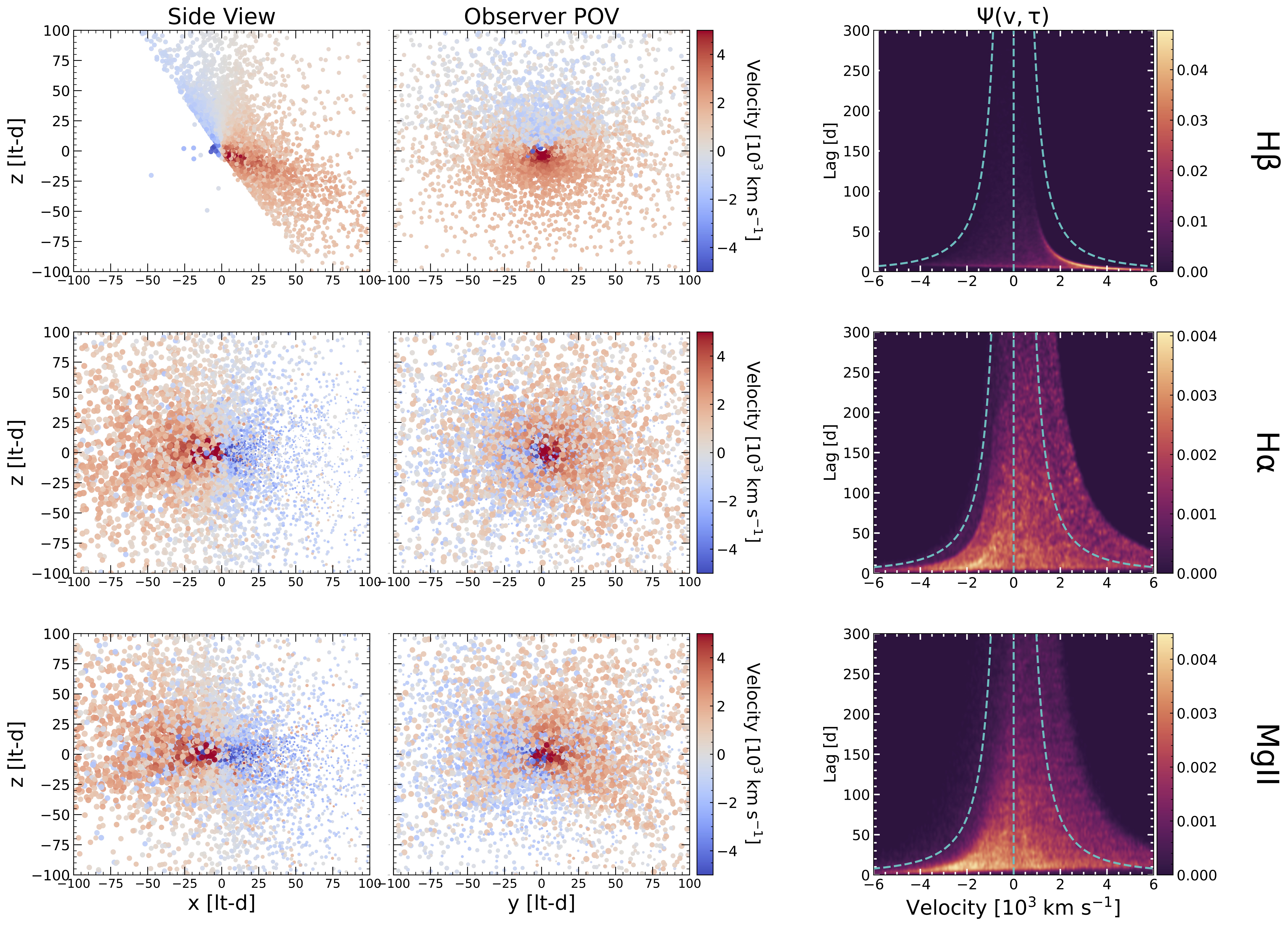}
    \caption{Same as Fig.~\ref{fig:brains_res_tot1} and Fig.~\ref{fig:brains_res_tot2}, but for the low state.}
    \label{fig:brains_res_low}
\end{figure*}

\begin{figure*}
    \centering
    \includegraphics[width=\textwidth]{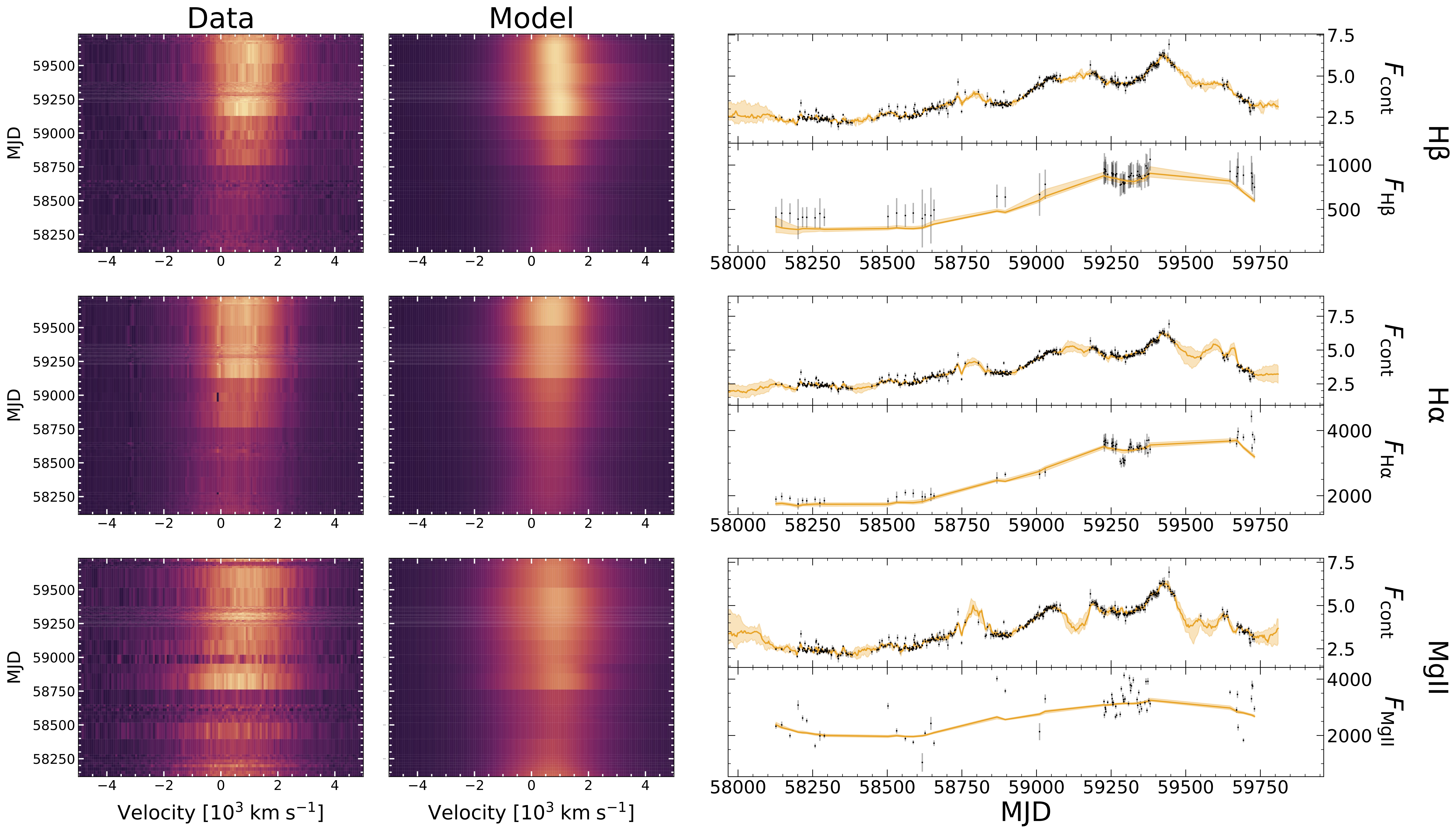}
    \includegraphics[width=\textwidth]{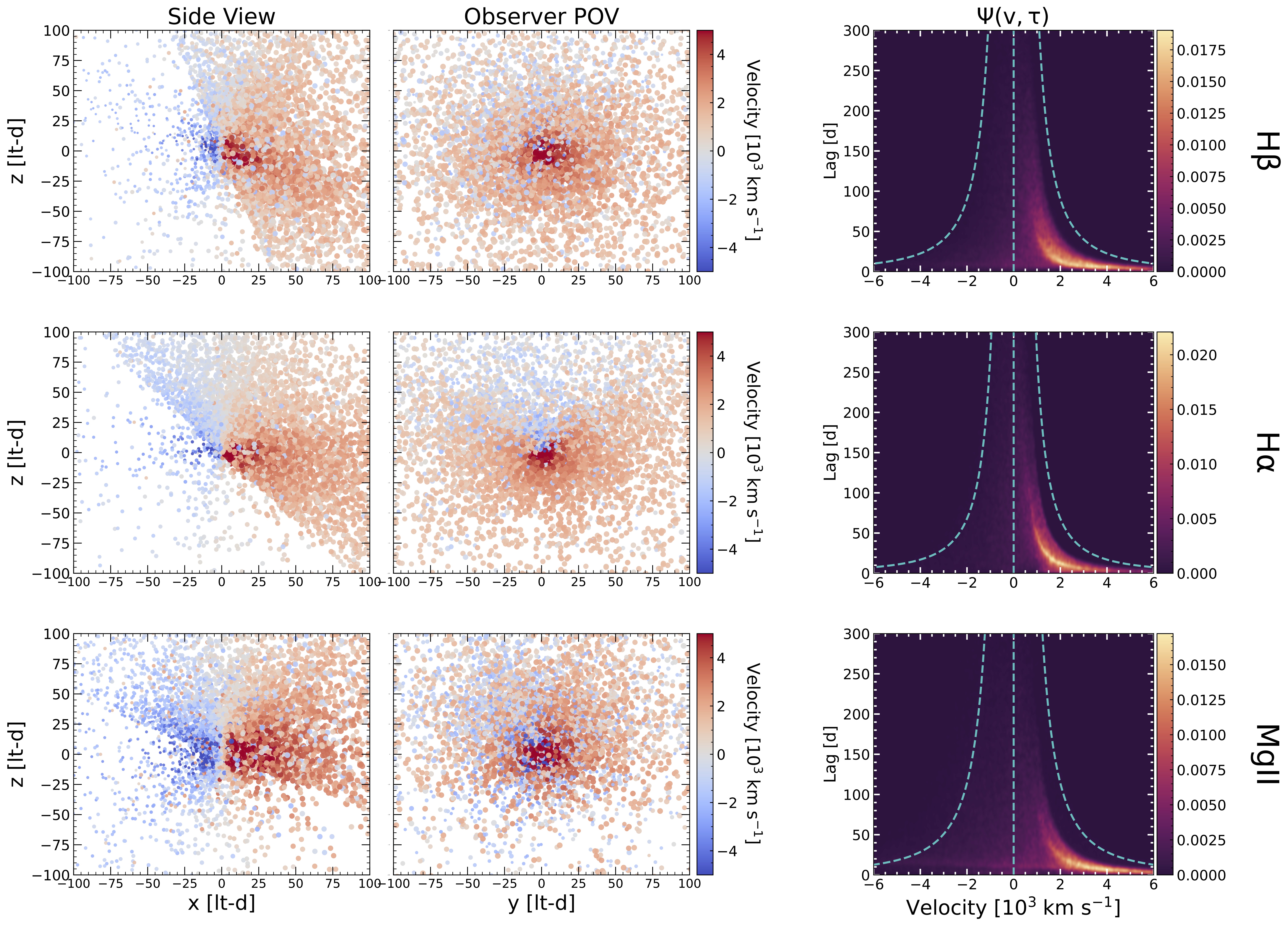}
    \caption{Same as Fig.~\ref{fig:brains_res_tot1} and Fig.~\ref{fig:brains_res_tot2}, but for the high state.}
    \label{fig:brains_res_high}
\end{figure*}

All best-fit model parameters for each \brains\, run, as well as the post-processing temperature $T$ used, are shown in Table.~\ref{tab:all_param}. Diagrams presenting the fits of the model to the data, as well as output cloud configurations and transfer functions $\Psi(v,\tau)$, are given in Figs~\ref{fig:brains_res_tot1}-\ref{fig:brains_res_high}.

\subsection{Model Interpretation}\label{subsec:model_interp}


\begin{table*}
\footnotesize
\begin{rotatetable*}
\caption{Best-fit Parameters For All {\tt BRAINS} Runs}
     \begin{adjustwidth}{-6cm}{}
\begin{tabular}{|l|l?c|c|c?c|c|c?c|c|c|}
\hline
\textbf{Parameter} & \textbf{Unit} & \multicolumn{3}{c?}{\textbf{Low State}} & \multicolumn{3}{c?}{\textbf{High State}} & \multicolumn{3}{c|}{\textbf{Full}} \\
\hline
\multicolumn{2}{|c?}{} & \hbeta & \halpha & \MgII & \hbeta & \halpha & \MgII & \hbeta & \halpha & \MgII \\
\hline
\setrow{\boldmath\bfseries}$\log_{10}(R_{BLR})$ & lt-day & $1.58^{+0.52}_{-0.00}$ & $2.04^{+0.00}_{-0.19}$ & $1.92^{+0.00}_{-0.00}$ & $2.03^{+0.02}_{-0.33}$ & $2.13^{+0.02}_{-0.20}$ & $1.94^{+0.32}_{-0.03}$ & $1.85^{+0.02}_{-0.14}$ & $2.39^{+0.00}_{-0.68}$ & $1.78^{+0.10}_{-0.11}$ \\[5pt]
$\beta$ &  & $1.51^{+0.07}_{-0.00}$ & $1.21^{+0.00}_{-0.16}$ & $1.20^{+0.00}_{-0.00}$ & $1.25^{+0.10}_{-0.06}$ & $1.02^{+0.07}_{-0.05}$ & $1.26^{+0.03}_{-0.33}$ & $1.22^{+0.09}_{-0.07}$ & $1.26^{+0.00}_{-0.19}$ & $1.07^{+0.07}_{-0.05}$ \\[5pt]
$F$ & $R_{BLR}$ & $0.14^{+0.00}_{-0.03}$ & $0.04^{+0.01}_{-0.03}$ & $0.07^{+0.00}_{-0.00}$ & $0.06^{+0.03}_{-0.03}$ & $0.02^{+0.03}_{-0.01}$ & $0.10^{+0.04}_{-0.08}$ & $0.03^{+0.01}_{-0.00}$ & $0.04^{+0.00}_{-0.00}$ & $0.05^{+0.03}_{-0.03}$ \\[5pt]
\setrow{\boldmath\bfseries}$i$ & deg & $30.12^{+59.77}_{-0.12}$ & $52.61^{+0.00}_{-26.76}$ & $56.58^{+0.00}_{-0.79}$ & $18.90^{+8.38}_{-3.40}$ & $41.46^{+37.35}_{-18.58}$ & $64.84^{+9.20}_{-47.42}$ & $48.39^{+20.11}_{-5.04}$ & $48.79^{+7.08}_{-0.00}$ & $76.88^{+8.70}_{-21.10}$ \\[5pt]
\setrow{\boldmath\bfseries}$\theta_{opn}$ & deg & $40.76^{+46.22}_{-0.10}$ & $66.63^{+0.00}_{-26.08}$ & $49.55^{+0.55}_{-0.00}$ & $53.77^{+10.30}_{-20.70}$ & $57.47^{+28.69}_{-11.66}$ & $74.88^{+7.30}_{-42.49}$ & $58.88^{+8.22}_{-3.41}$ & $49.45^{+0.00}_{-0.00}$ & $80.58^{+4.85}_{-29.44}$ \\[5pt]
$\kappa$ &  & $-0.15^{+0.65}_{-0.01}$ & $-0.49^{+0.77}_{-0.00}$ & $-0.44^{+0.00}_{-0.02}$ & $0.40^{+0.08}_{-0.21}$ & $0.27^{+0.01}_{-0.76}$ & $0.32^{+0.04}_{-0.02}$ & $0.38^{+0.05}_{-0.86}$ & $0.45^{+0.00}_{-0.90}$ & $-0.41^{+0.06}_{-0.07}$ \\[5pt]
$\gamma$ &  & $2.31^{+1.93}_{-0.20}$ & $2.15^{+2.02}_{-0.20}$ & $4.49^{+0.24}_{-0.57}$ & $2.65^{+0.83}_{-1.41}$ & $2.39^{+2.30}_{-0.92}$ & $4.35^{+0.35}_{-0.60}$ & $2.80^{+1.81}_{-0.69}$ & $3.10^{+0.00}_{-0.38}$ & $2.11^{+1.49}_{-0.62}$ \\[5pt]
$\xi$ &  & $0.01^{+0.89}_{-0.00}$ & $0.97^{+0.00}_{-0.57}$ & $0.93^{+0.03}_{-0.00}$ & $0.19^{+0.12}_{-0.08}$ & $0.11^{+0.58}_{-0.04}$ & $0.30^{+0.09}_{-0.15}$ & $0.52^{+0.31}_{-0.19}$ & $0.97^{+0.00}_{-0.10}$ & $0.68^{+0.19}_{-0.27}$ \\[5pt]
\setrow{\boldmath\bfseries}$\log_{10}(M_{BH})$ & $M_{\odot}$ & $7.65^{+0.27}_{-0.01}$ & $7.73^{+0.15}_{-0.04}$ & $7.74^{+0.00}_{-0.00}$ & $7.85^{+0.15}_{-0.37}$ & $7.72^{+0.40}_{-0.02}$ & $7.95^{+0.08}_{-0.04}$ & $7.66^{+0.05}_{-0.13}$ & $7.89^{+0.00}_{-0.51}$ & $7.63^{+0.09}_{-0.09}$ \\[5pt]
\hline
\setrow{\boldmath\bfseries}$f_{ellip}$ &  & $0.00^{+0.07}_{-0.00}$ & $0.20^{+0.08}_{-0.00}$ & $0.34^{+0.00}_{-0.00}$ & $0.13^{+0.18}_{-0.13}$ & $0.04^{+0.20}_{-0.04}$ & $0.19^{+0.35}_{-0.01}$ & $0.04^{+0.02}_{-0.03}$ & $0.07^{+0.00}_{-0.06}$ & $0.04^{+0.14}_{-0.03}$ \\[5pt]
\setrow{\boldmath\bfseries}$f_{flow}$ &  & $0.27^{+0.10}_{-0.17}$ & $0.62^{+0.00}_{-0.39}$ & $0.98^{+0.00}_{-0.00}$ & $0.28^{+0.08}_{-0.19}$ & $0.27^{+0.62}_{-0.17}$ & $0.14^{+0.32}_{-0.05}$ & $0.26^{+0.30}_{-0.19}$ & $0.38^{+0.24}_{-0.00}$ & $0.73^{+0.20}_{-0.22}$ \\[5pt]
$\log_{10}(\sigma_{\rho, circ})$ & $v_{circ}$ & $-2.58^{+0.99}_{-0.17}$ & $-1.08^{+0.00}_{-1.08}$ & $-2.36^{+0.24}_{-0.00}$ & $-2.10^{+0.24}_{-0.44}$ & $-2.07^{+0.37}_{-0.59}$ & $-2.41^{+0.63}_{-0.13}$ & $-1.73^{+0.57}_{-0.70}$ & $-1.05^{+0.00}_{-1.12}$ & $-2.06^{+0.73}_{-0.68}$ \\[5pt]
$\log_{10}(\sigma_{\theta, circ})$ & $v_{circ}$ & $-2.11^{+1.58}_{-0.37}$ & $-1.96^{+0.04}_{-0.98}$ & $-0.72^{+0.00}_{-0.19}$ & $-2.02^{+1.07}_{-0.95}$ & $-1.84^{+1.51}_{-1.08}$ & $-2.06^{+2.02}_{-0.45}$ & $-2.40^{+1.41}_{-0.36}$ & $-1.93^{+0.90}_{-0.00}$ & $-1.03^{+0.78}_{-1.22}$ \\[5pt]
$\log_{10}(\sigma_{\rho, rad})$ & $v_{circ}$ & $-2.20^{+0.99}_{-0.63}$ & $-1.94^{+0.70}_{-0.00}$ & $-1.17^{+0.00}_{-0.00}$ & $-1.63^{+0.50}_{-0.50}$ & $-1.90^{+0.82}_{-0.68}$ & $-2.00^{+0.84}_{-0.64}$ & $-2.27^{+0.96}_{-0.32}$ & $-1.44^{+0.28}_{-0.00}$ & $-1.82^{+0.63}_{-0.80}$ \\[5pt]
$\log_{10}(\sigma_{\theta, rad})$ & $v_{circ}$ & $-2.22^{+0.09}_{-0.48}$ & $-2.12^{+0.00}_{-0.46}$ & $-1.02^{+0.00}_{-0.11}$ & $-1.86^{+0.68}_{-0.67}$ & $-2.44^{+0.87}_{-0.33}$ & $-2.07^{+0.72}_{-0.55}$ & $-1.48^{+0.72}_{-1.25}$ & $-1.95^{+0.31}_{-0.00}$ & $-1.61^{+0.63}_{-0.82}$ \\[5pt]
$\theta_{e}$ & deg & $2.29^{+1.62}_{-0.61}$ & $24.68^{+0.00}_{-24.54}$ & $15.54^{+3.99}_{-0.00}$ & $25.84^{+5.83}_{-9.42}$ & $16.65^{+26.70}_{-15.18}$ & $14.44^{+33.35}_{-10.87}$ & $7.36^{+27.87}_{-4.98}$ & $3.53^{+0.00}_{-0.59}$ & $15.96^{+11.64}_{-11.96}$ \\[5pt]
$\log_{10}(\sigma_{turb})$ & $v_{circ}$ & $-2.65^{+0.70}_{-0.67}$ & $-2.43^{+0.00}_{-0.19}$ & $-3.17^{+0.22}_{-0.00}$ & $-1.97^{+0.88}_{-1.26}$ & $-2.53^{+0.84}_{-0.59}$ & $-2.24^{+1.21}_{-0.89}$ & $-2.50^{+0.88}_{-1.02}$ & $-2.96^{+0.00}_{-0.55}$ & $-2.17^{+0.96}_{-1.16}$ \\[5pt]
\hline
$\Delta V_{line}$ & $\rm km \; s^{-1}$ & $323.81^{+86.62}_{-66.05}$ & $466.24^{+24.56}_{-158.93}$ & $540.04^{+21.16}_{-23.29}$ & $268.87^{+57.32}_{-40.80}$ & $346.38^{+22.72}_{-115.38}$ & $304.54^{+67.98}_{-25.79}$ & $227.20^{+49.16}_{-53.76}$ & $384.71^{+36.76}_{-127.07}$ & $253.77^{+54.97}_{-49.85}$ \\[5pt]
$\sigma_{sys, line}$ & $\rm f_\lambda$ Units & $0.11^{+0.01}_{-0.01}$ & $0.29^{+0.01}_{-0.01}$ & $0.92^{+0.02}_{-0.03}$ & $0.26^{+0.06}_{-0.01}$ & $0.67^{+0.02}_{-0.03}$ & $1.78^{+0.08}_{-0.06}$ & $0.19^{+0.01}_{-0.01}$ & $0.65^{+0.03}_{-0.01}$ & $1.76^{+0.14}_{-0.14}$ \\[5pt]
$\sigma_{sys, con}$ & $\rm f_\lambda$ Units & \nodata & \nodata & \nodata & \nodata & \nodata & \nodata & \nodata & \nodata & \nodata \\[5pt]
$\log_{10}( \sigma_d )$ & $\rm f_\lambda$ Units & $-1.83^{+0.03}_{-0.04}$ & $-1.76^{+0.00}_{-0.00}$ & $-1.79^{+0.00}_{-0.00}$ & $-1.91^{+0.03}_{-0.04}$ & $-1.78^{+0.05}_{-0.05}$ & $-1.73^{+0.01}_{-0.04}$ & $-1.98^{+0.03}_{-0.02}$ & $-1.93^{+0.02}_{-0.02}$ & $-1.97^{+0.02}_{-0.03}$ \\[5pt]
$\log_{10}( \tau_d )$ & d & $2.75^{+0.23}_{-0.22}$ & $2.77^{+0.22}_{-0.22}$ & $2.76^{+0.21}_{-0.22}$ & $2.71^{+0.09}_{-0.11}$ & $2.83^{+0.16}_{-0.20}$ & $2.70^{+0.18}_{-0.08}$ & $2.97^{+0.19}_{-0.22}$ & $2.99^{+0.25}_{-0.20}$ & $3.04^{+0.11}_{-0.19}$ \\[5pt]
B & $\rm d^{-1}$ & $-0.34^{+1.02}_{-0.91}$ & $-0.25^{+1.14}_{-0.97}$ & $-0.16^{+0.98}_{-1.09}$ & $-0.53^{+1.67}_{-0.30}$ & $0.77^{+0.73}_{-1.45}$ & $0.57^{+0.96}_{-0.84}$ & $-0.04^{+0.96}_{-1.18}$ & $0.19^{+1.43}_{-1.33}$ & $-0.18^{+0.70}_{-0.81}$ \\[5pt]
$A$ &  & $-0.00^{+0.03}_{-0.01}$ & $0.05^{+0.02}_{-0.04}$ & $-0.13^{+0.01}_{-0.01}$ & $-0.29^{+0.04}_{-0.02}$ & $-0.11^{+0.03}_{-0.02}$ & $-0.10^{+0.00}_{-0.01}$ & $-0.02^{+0.02}_{-0.02}$ & $0.00^{+0.01}_{-0.01}$ & $-0.02^{+0.01}_{-0.02}$ \\[5pt]
$A_g$ &  & $0.29^{+0.08}_{-0.14}$ & $0.31^{+0.06}_{-0.06}$ & $1.64^{+0.14}_{-0.30}$ & $0.76^{+0.07}_{-0.12}$ & $0.08^{+0.03}_{-0.12}$ & $-0.20^{+0.03}_{-0.04}$ & $0.39^{+0.03}_{-0.05}$ & $0.22^{+0.03}_{-0.04}$ & $-0.11^{+0.05}_{-0.04}$ \\[5pt]
\hline
$T$ & & 5.00 & 2.60 & 2.50 & 131.70 & 15.80 & 4.85 & 48.69 & 78.56 & 101.00 \\[5pt]
\hline
\end{tabular}
\label{tab:all_param}
    \end{adjustwidth}
\end{rotatetable*}
\end{table*}


\subsubsection{Model Geometry}\label{subsubsec:geometry}

All best-fit models represent a thick disk BLR moderately edge-on along the LOS of the observer. Using the joint posteriors for $M_{\rm BH}$, the best-fit values are $\log_{10}(M_{\rm BH} / M_{\odot}) = $ $7.70^{+0.16}_{-0.04}$ (low state), $7.76^{+0.30}_{-0.16}$ (high state), and $7.66^{+0.12}_{-0.13}$ (full state). 
As the estimates for $M_{\rm BH}$ for each model are independent, the fact that the best estimate for $M_{\rm BH}$ is consistent across states displays the consistency in modeling across \brains\, runs. The joint posteriors across lines yield best estimates of $i = 27.94^{\circ} \,{}^{+6.80}_{-5.07}$ (low state), $53.53^{\circ} \,{}^{+16.81}_{-23.74}$ (high state), $53.29^{\circ} \,{}^{+7.29}_{-6.55}$ (full state) and $\theta_{\rm opn} = 41.51^{\circ} \,{}^{+4.86}_{-4.62}$ (low state), $56.32^{\circ} \,{}^{+20.53}_{-12.63}$ (high state), $54.86^{\circ} \,{}^{+5.83}_{-4.74}$ (full state). The high and full states have consistent inclinations, while the low state inclination is significantly lower (i.e., more face-on), primarily from the low state of \hbeta. The inclination of the system across lines and states is $i \approx 50^{\circ}$, more edge-on than most prior dynamical modeling results \citep[e.g.,][]{GrierEtAl2017c, VillafanaEtAl2022a}. The disk opening angle is consistent across states. For most lines and states $i \sim \theta_{\rm opn}$, with our line of sight skimming the upper edge of the BLR disk. Generally the sizes of the BLRs appear in the following order: $R_{\rm H\beta} \lesssim R_{\rm MgII } \lesssim R_{\rm H\alpha}$ (see Section \ref{subsec:stratification}), although the uncertainties in these radii can be large. The shapes of the radial distribution for all lines in all states are consistent with one another, having heavier tails than exponential distributions ($\beta \approx 1.2$). Comparing the BLR geometry across lines, the inclination for \MgII\, is generally higher (i.e., edge-on) than for the Balmer lines, and \hbeta\, is the least inclined (i.e., face-on). \citet{WilliamsEtAl2020a} find the BLR for HILs have a higher inclination than the more distant LIL BLRs. This could suggest a flared disk with a changing scale height, or a more complex geometry not accounted for in the model.

The results from each run for the asymmetry parameters ($\kappa, \gamma, \xi$) are less constrained, fluctuate between runs, and display significant multimodality. The opacity of the midplane varies from completely opaque ($\xi = 0$) to completely transparent ($\xi=1$) across runs. Some runs display far-side preference ($\kappa \approx 0.5$), while others display near-side preference ($\kappa \approx -0.5$), owing to a degeneracy with the binary inflow/outflow parameter ($f_{\rm flow}$) (see Section~\ref{subsubsec:kinematics}). The disk outer face clustering parameter ($\gamma$) is the same across states for \hbeta\, and \halpha, with moderate clustering. \MgII\, is less constrained, with the low and high states showing a preference for clustering, and the full state showing moderate clustering.

\subsubsection{Model Kinematics}\label{subsubsec:kinematics}

All runs indicate that most BLR cloud orbits are not in bound elliptical orbits, but rather inflowing or outflowing. Each line shows a varying amount of clouds on elliptical orbits, with the \hbeta\, BLR having almost no clouds on elliptical orbits, and the \MgII\, BLR having just over 30\% of clouds on elliptical orbits in the low state. On average, for all lines and all states, $f_{\rm ellip} \lesssim 0.3$. All models have $\theta_e \in [0,30]^{\circ}$, where $\theta_e$ is the location of the center of the Gaussian distribution of cloud velocities in radial velocity $(v_r)$ - tangential velocity $(v_\phi)$ space. $\theta_e \sim 0^{\circ}$ represents velocities centered around the escape velocity, whereas $\theta_e \sim 45^{\circ}$ represents highly eccentric (i.e., ``bound inflowing/outflowing'') orbits (see \citetalias{PancoastEtAl2014b}). The impact of macroturbulent motions is minimal in the BLR for all models, with $\sigma_{\rm turb} \approx 0.001 - 0.01$.

This lack of viriality seemingly proves contrary to the transfer functions produced from each model, with most of the transfer function lying within the ``virial envelope''.  However, while the orbit of many BLR clouds in the model may be unbound, the dynamics of the BLR as a whole are close to a Keplerian scenario. The values for $\theta_e$ can vary across models, with the high state for \hbeta\, having $\theta_e \approx 26^{\circ}$ for example. This suggests that a fraction of clouds will be on highly eccentric bound orbits. This explains why the intensity of the transfer functions mostly lie within the virial envelope, though asymmetrically. A BLR with many highly eccentric ``inflowing'' Keplerian orbits would produce a transfer function with most intensity at positive velocities, while still remaining within the virial envelope. Additionally, many of the unbound orbits are close to Keplerian velocity, contributing to observed flux on the edge of the virial envelope, validated by the small widths of the velocity distributions (Table~\ref{tab:all_param}). It should be noted that even though the BLR is somewhat non-virial, this does not affect the measurement of $M_{\rm BH}$, as in velocity-resolved RM. The dynamical model makes no assumption of viriality, only assuming that the clouds in the BLR act under the gravitational influence of the SMBH.



While consistent values across runs for $f_{\rm ellip}$ place most clouds in the BLR on non-elliptical orbits, the distinction between outflow and inflow with the binary $f_{\rm flow}$ parameter is less constrained within individual runs, and across lines and states. The large line center shifts toward higher wavelengths in all lines suggest inflow within the BLR, as suggested by \citetalias{FriesEtAl2023}. However, there is a degeneracy between $f_{\rm flow}$ and the asymmetry parameters $\kappa$ and $\xi$. $\kappa$ assigns radiative weights to the clouds, depending on whether they are located on the far- or near-side of the disk in relation to the observer (i.e., if the clouds prefer to radiate back towards the photoionizing source or not). In an inflow scenario, a large amount of red flux can be reproduced in the model if the near-side of the disk is weighted higher. The same amount of red flux can be reproduced in an outflow scenario if the far-side is weighted higher, a scenario seen in photoionization modeling \citep{FerlandEtAl1992}. An increase in opacity of the midplane from $\xi$ can also increase the amount of red flux if the BLR prefers the radiation from the near-side in an inflow scenario. This degeneracy and the width and multimodality of the $\kappa$, $\xi$, and $f_{\rm flow}$ posteriors suggest the direction of the orbits is relatively unconstrained across runs; for one run a far-side preference and outflow may fit the data better, but for another run the opposite scenario may fit better.

\MgII\, shows best-fit $f_{\rm flow}$ values representative of outflow in the low and full states, with $\kappa$ representative of a far-side preference. However, due to the low response of the \MgII\, BLR to the continuum and longer lags, many combinations of parameters can reproduce the observed line profiles. It should be noted that all model results for \hbeta\, firmly indicate inflow. While it has been suggested that blueshifted broad emission from \MgII\, is indicative of an outflow with preferential emission from the far-side of the disk \citep{MarzianiEtAl2013, MarzianiEtAl2013a}, the empirical analysis from \citetalias{FriesEtAl2023} and the presence of inflow for \hbeta\, point toward an unbroken degeneracy. However, as suggested in \citet{BentzEtAl2021a}, differences in kinematics across lines can occur due to a form of transient ``weather'' or photoionization physics.

\citetalias{FriesEtAl2023} provide an empirical model given their analysis of RM160's emission line properties, suggesting an inflow scenario with a gradient in the inflow velocity decreasing outwardly, and a ``hotspot'' (i.e., an azimuthal asymmetry) moving rapidly around the BLR during the monitoring period. We confirm strong inflow closer to the SMBH for both the low and high state, combining the population of clouds from all three lines. While the agreement with a gradient of radial velocity in both the near-side inflow and far-side outflow scenarios is validating, the dynamical model presented in \citetalias{PancoastEtAl2014b} is not able to accurately capture the complex dynamics of the hotspot in \citetalias{FriesEtAl2023}'s model.

\subsubsection{Inferences Across Best-Fit Models}\label{subsubsec:joint}

The geometries are similar across states for all lines except $R_{\rm BLR}$, similar to \citet{PancoastEtAl2018} who find that the \hbeta\, $R_{\rm BLR}$ almost doubles in Arp 151 from 2008 to 2011. The near-side/far-side preference parameter $(\kappa)$ changes drastically for \MgII\, from the low state to the high state, but so does $f_{\rm flow}$, suggestive of the degeneracy and lack of constraint on the radial direction parameter. $f_{\rm ellip}$ also decreases from the low state to the high state for both \halpha\, (marginally) and \MgII\, (somewhat significantly), although the general state of the kinematics remains the same. Potential changes in the inflow/outflow fraction could imply the existence of transient ``weather'' occurring within the BLR, changing kinematic conditions throughout the monitoring period. Transient events have been reported in NGC 5548, displaying a ``BLR holiday'' wherein the continuum and emission line variations became decorrelated \citep{GoadEtAl2016}. This behavior has been attributed to a change in covering factor, a change in density of a disk wind, or a quickly orbiting azimuthal asymmetry \citep{DehghanianEtAl2019, HorneEtAl2021}. While these potential explanations are not explicitly modeled by the dynamical model, the consequent changes in geometry and kinematics would lead to changes in the best-fit model parameters. The linearity of the response $A_g$ remains similar across runs, except for the low state of \MgII, which is significantly higher. 

\begin{figure*}
    \centering
    \includegraphics[width=\textwidth]{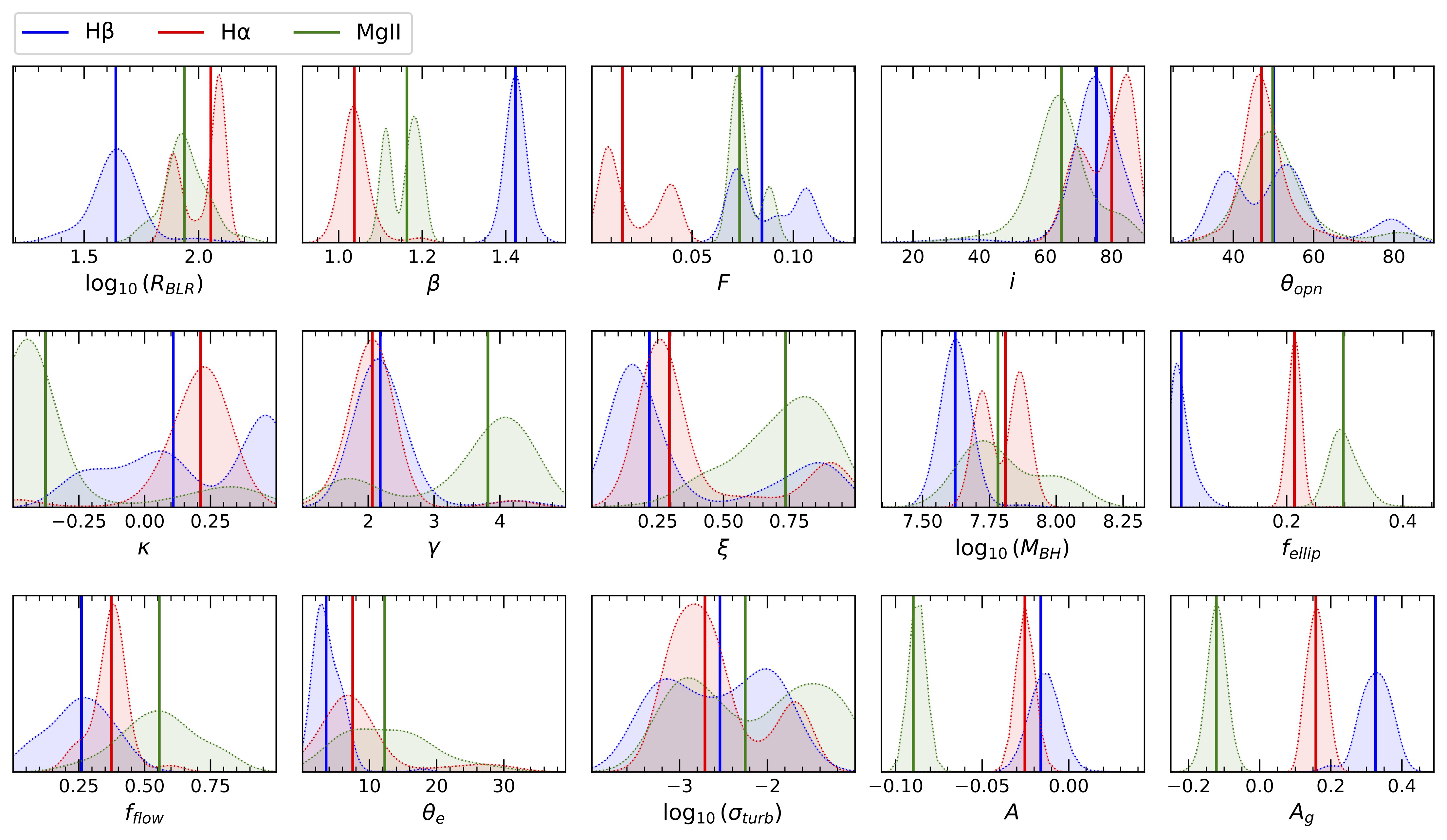}
    \caption{Joint posteriors for a given line combining posteriors from each state. Individual likelihood functions were obtained using Gaussian KDEs. Median values for each line's posterior are marked using colored solid vertical lines.}
    \label{fig:posterior_comp_statecombined}
\end{figure*}

We follow the procedure described in \citet{PancoastEtAl2018} to create joint posteriors across states for each line (Fig.~\ref{fig:posterior_comp_statecombined}). In short, the unweighted likelihood distributions for each state are multiplied together, then multiplied by a given parameter's prior, to produce a joint posterior. While this approach is not applicable or useful for parameters that change significantly from state to state (e.g., $R_{\rm BLR}, F$), we include them for the sake of completeness, and they can be identified easily by inspecting for multiple peaks in the posterior. Using this approach verifies that the inclination and opening angles of the disk are consistent across lines, without any weighting. $M_{\rm BH}$ is also consistent across lines. While $R_{\rm BLR}$ may change across states, we see the same order of stratification in the joint posteriors as in the $M_{\rm BH}$-weighted posteriors. However, this process widens the posteriors, causing all three BLR sizes to overlap in uncertainty. In particular, the size of the \halpha\, region is drawn closer to that of \MgII. While the \halpha\, posterior is wide and multimodal, a large portion of it is located near the calculated dust sublimation radius of $\sim 128$ light-days, suggesting that the \halpha\, region of the BLR is located near the inner edge of the torus. $f_{\rm flow}$ is more unconstrained for \MgII\, spanning almost the entire prior, while \hbeta\, and \halpha\, are firmly inflow. All lines show either mostly inflowing or a combination of inflowing and elliptical orbits, with the amount of relative inflow decreasing from \hbeta\, to \halpha\, to \MgII, suggestive of a radial velocity gradient.  

\subsection{Velocity-Resolved Lags}\label{subsec:vrrm}

\begin{figure*}
    \centering
    \includegraphics[width=\textwidth]{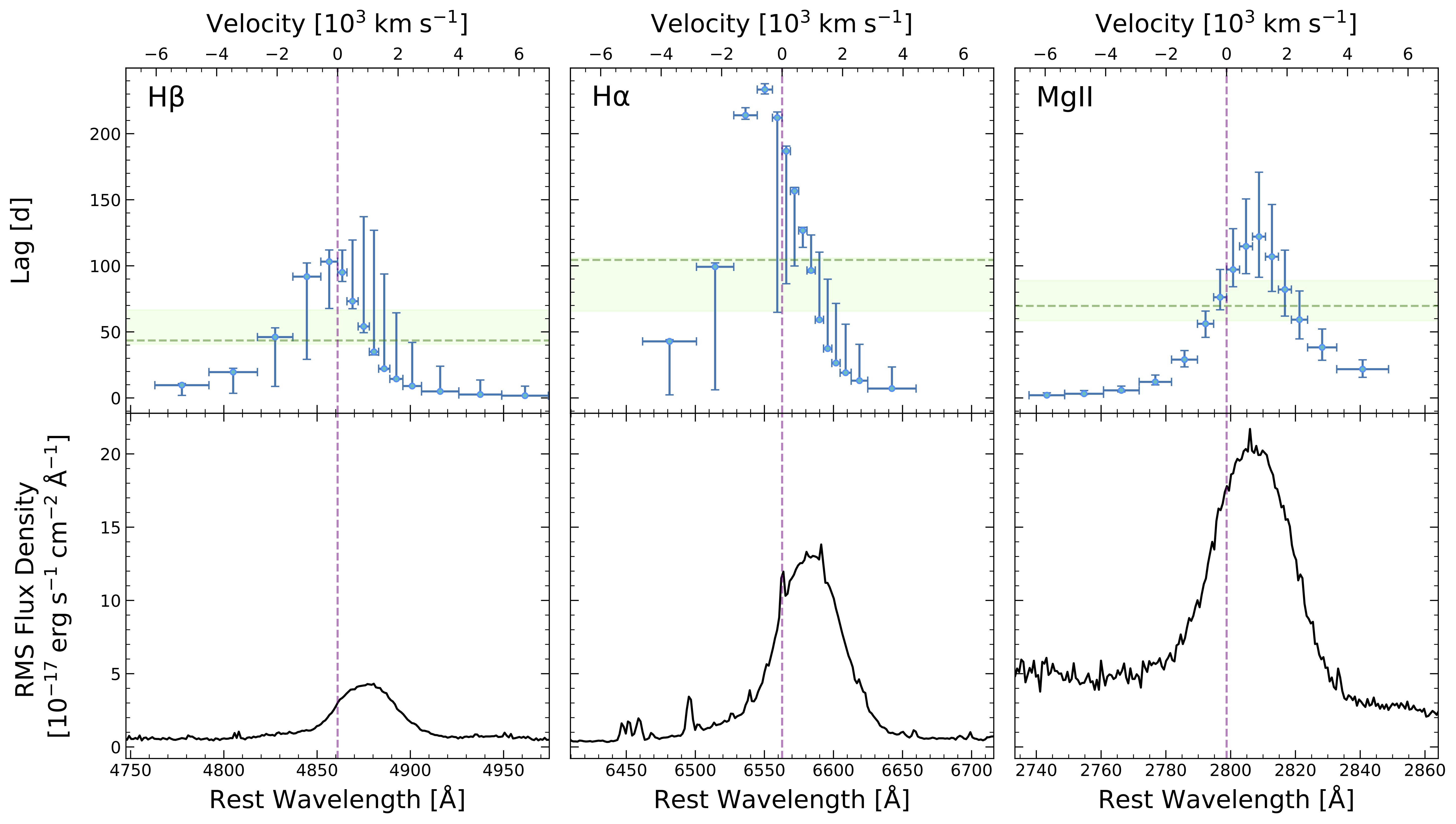}
    \caption{\emph{Top:} The distribution of rest-frame lags output from the model as a function of wavelength for each of the three lines in the full state. The green dotted horizontal line represents the lag over the entire wavelength range, with the $16^{th} - 84^{th}$ percentile range shaded in light green. The vertical dashed purple line indicates the central wavelength of the line. \emph{Bottom:} The RMS line profile for each of the three lines in the full state. }
    \label{fig:lagplot_tot}
\end{figure*}

\begin{figure*}
    \centering
    \includegraphics[width=\textwidth]{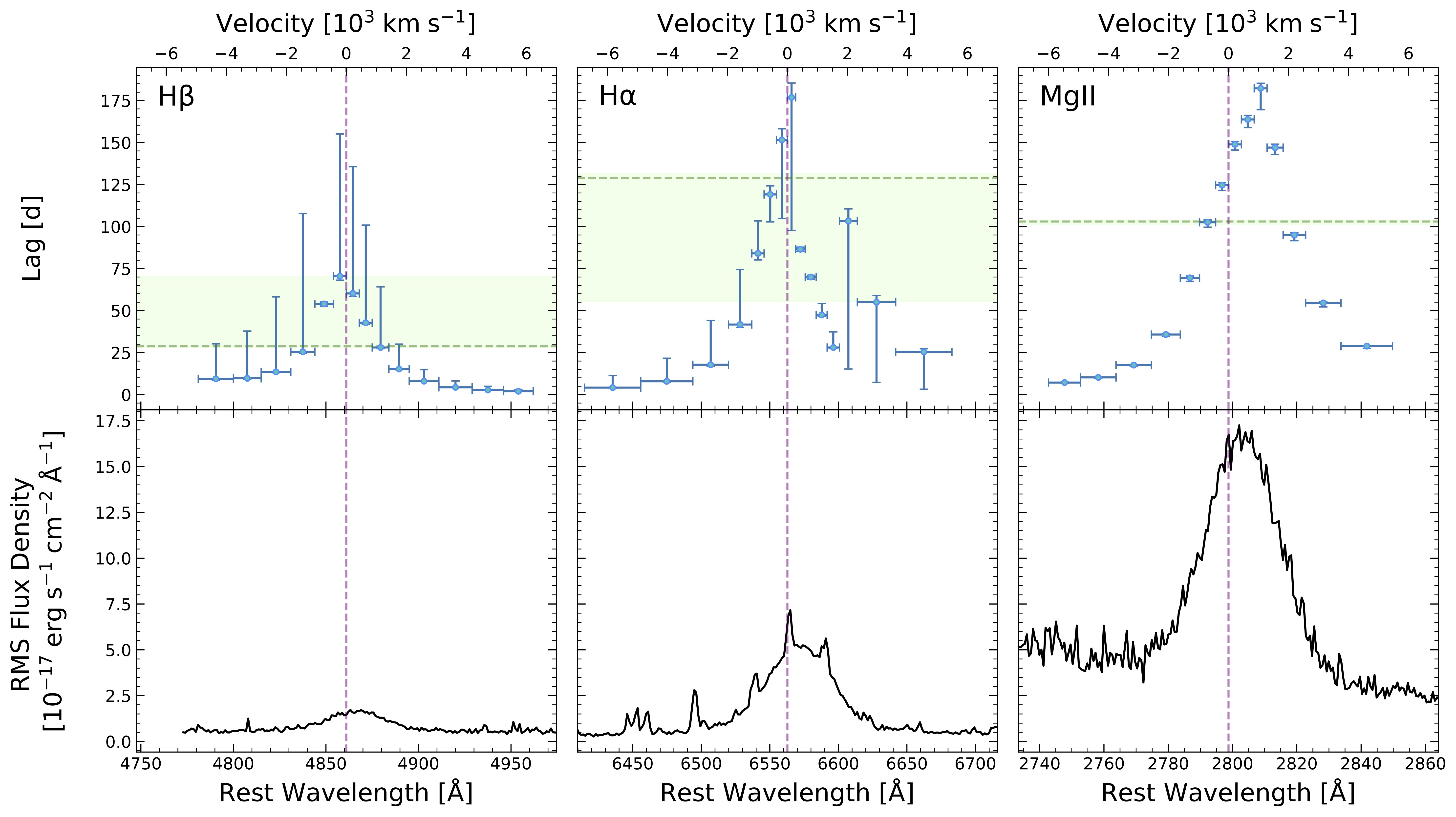}
    \caption{Same as Fig.~\ref{fig:lagplot_tot}, but for the low state.}
    \label{fig:lagplot_low}
\end{figure*}

\begin{figure*}
    \centering
    \includegraphics[width=\textwidth]{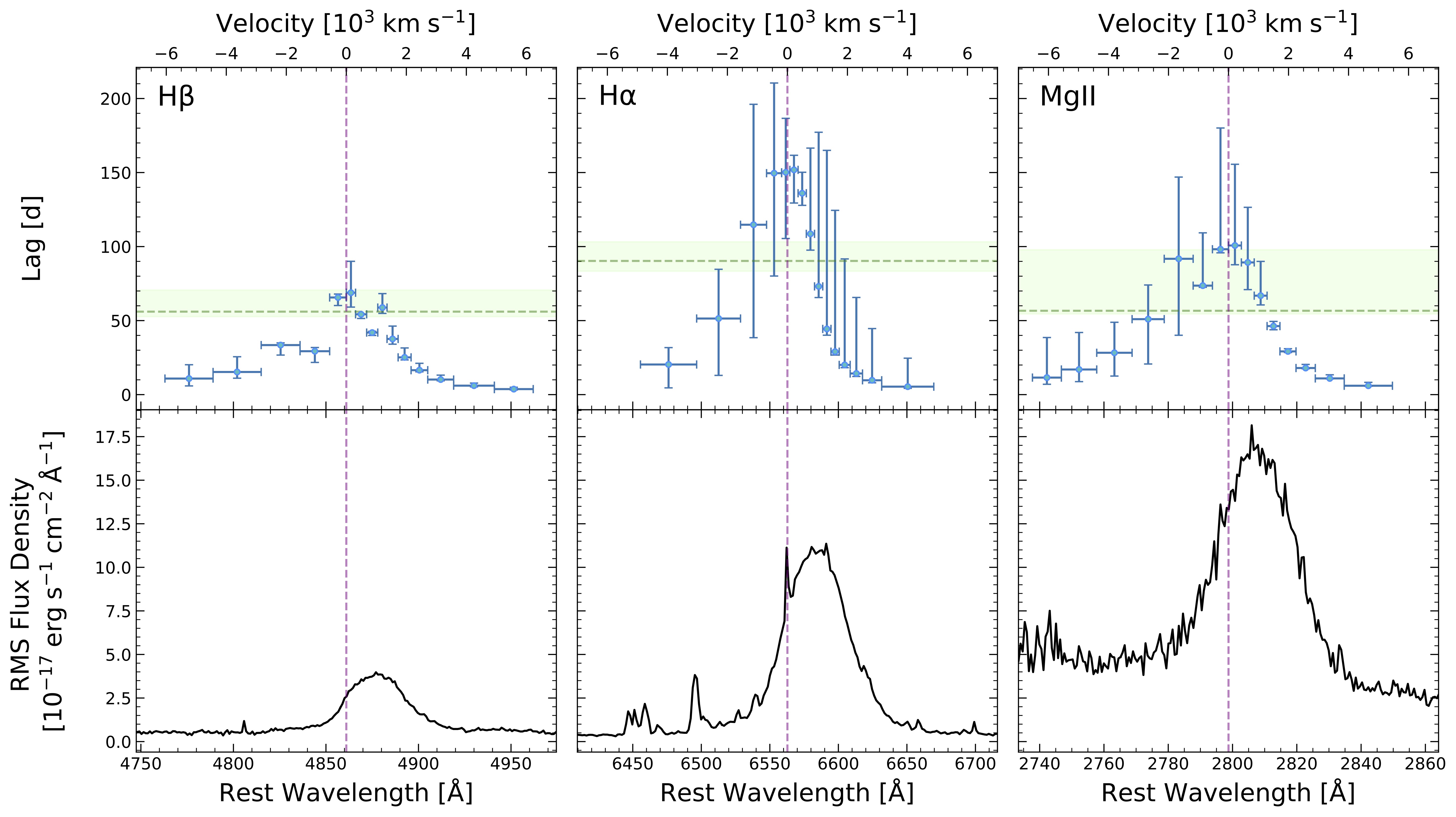}
    \caption{Same as Fig.~\ref{fig:lagplot_tot}, but for the high state.}
    \label{fig:lagplot_high}
\end{figure*}

By integrating $\Psi(v, \tau)$ for a given range of velocities, we obtain model-produced, velocity-resolved lags for a given line and a given state. We follow the same procedure as in \citet{DenneyEtAl2009} and \citet{GrierEtAl2013} to create velocity-binned lags using the root-mean-square (RMS) profile of a given line. We separate each RMS profile into 15 bins of equal flux, centered around the rest-wavelength of the line (i.e., $v=0$), extending to velocities of $v = \pm 7000$ km s$^{-1}$. Our velocity-resolved lags, along with the RMS profiles, are shown in Figs.~\ref{fig:lagplot_tot}-\ref{fig:lagplot_high}. 

The interpretation of these lag structures are consistent with those from Section~\ref{subsec:model_interp}. However, the uncertainty in the individual velocity-resolved lags leaves room for multiple interpretations. For example, the lag profile for \halpha\, in the full state is consistent with peaking in the red-wing, center, and blue-wing of the line profile. Generally, all lag profiles peak near the center, and fall towards the edges of the line, representative of Keplerian motion. Profiles peaking in the blue wing of the line represent inflow (i.e., short lags are redshifted), while profiles peaking in the red wing of the line represent outflow (i.e., short lags are blueshifted). A purely inflowing BLR would display large lags in the blue wing and small lags in the red wing, not dropping off to zero on the blue side. The fact that we see a drop-off on both sides of the emission line suggests that while a significant fraction of BLR clouds are on truly unbound orbits, they move close to the Keplerian velocity (see Section~\ref{subsubsec:kinematics}).

Furthermore, it is difficult to determine the size of the BLR from the lag alone, as the inclination and other asymmetric properties of the geometry can cause over- or underestimation \citep{McLureDunlop2002, DecarliEtAl2008, RunnoeEtAl2013, Shen2013}. It is even more difficult to recover the kinematic characteristics of the BLR, including the amount of inflow or the gradient of the BLR radial velocity profile. Lag profiles such as that for \hbeta\, in the low state with almost no elliptical orbits are similar to the lag profile for \halpha\, in the low state with 20\% elliptical orbits. These profiles display the difficulty in interpreting velocity-resolved RM results in relation to a given model, as even when the model is inherently tied to the lag results, the lag profiles are still somewhat noisy and the interpretation can become muddled. 

Generally, we produce similar lag profiles from the low state to the high state -- they appear to be dominated by virial motion with some influence of inflowing/outflowing cloud orbits. Fries et al.~(2024, in prep.) have performed velocity-resolved RM using data from RM160, albeit with slightly different definitions of the low and high states. For \hbeta, they report large lags in the blue wing of the profile and short lags in the red wing (i.e., pure inflow), a more extreme scenario than presented in our best-fit models. They find that the kinematic structure of the \halpha\, BLR changes drastically, showing a virialized symmetric profile in the low state and a purely inflowing profile in the high state. While the interpretation obtained from velocity-resolved RM can be unclear from uncertainties in the lags, such a difference in results between velocity-resolved RM and dynamical modeling suggests that these methods can lead to different conclusions. This differs from \citet{VillafanaEtAl2022a} whose dynamical modeling results agree with lag profiles constructed via velocity-resolved RM. More comparisons will need to be done between the two methods to verify which more accurately traces the kinematics of the BLR.

\subsection{BLR Stratification}\label{subsec:stratification}

Photoionization models predict different sizes and lags for the BLR, depending on the emission line studied. Prior work on photoionization modeling predicts that LILs have larger and more extended BLRs than HILs \citep{ClavelEtAl1991, GoadEtAl1993}. However, it has been difficult to study the \MgII\, BLR with theoretical modeling and data-driven methods such as RM. There have been a plethora of mixed results analyzing the \MgII\, BLR in the past. \MgII\, reverberates alongside the Balmer lines, although these two lines differ in terms of their responses to the continuum \citep[e.g.,][]{SunEtAl2015,GuoEtAl2020}, their breathing properties \citep{WangEtAl2020, GuoEtAl2020}, and their intrinsic $R-L$ relations observed by RM \citep{GuoEtAl2020, YangEtAl2020a}. Many studies report that the FWHM of \hbeta\, and \MgII\, positively correlate with each other \citep[e.g.,][]{ShenEtAl2008, WangEtAl2009, Kovacevic-DojcinovicPopovic2015}, though radiative transfer from the Balmer lines and continuum may influence the scenarios in which \MgII\, is emitted \citep{KoristaGoad2000}. In addition, due to \MgII\, having a significant fraction of its emission dominated by collisional excitation, it is more sensitive to environmental conditions (e.g., the ionization parameter) than the recombination-dominated Balmer lines \citep{KoristaGoad2004}.

Prior works have demonstrated that the sizes of the BLRs of these different lines display different lags and sizes (both apparent and physical), due to their different responsivities. It is well-known that higher-order Balmer lines (e.g., \hbeta) are more variable (hence more responsive) than lower-order Balmer lines (e.g., \halpha), and lie closer to the SMBH \citep{KoristaGoad2004}. Photoionization models find that \MgII\, lies farther out than the Balmer lines, and has a more widely distributed BLR \citep{OBrienEtAl1995, KoristaGoad2000, GuoEtAl2020}. However, this can change when considering different optical depths, and the ranges for the mean emissivity-weighted radius can change based on different physical conditions within the BLR \citep{OBrienEtAl1995}. Furthermore, recent RM studies find that lags obtained from \MgII\, are marginally larger than \citep{ShenEtAl2024} or similar to \citep{HomayouniEtAl2020, YuEtAl2021} those from \hbeta. \citetalias{FriesEtAl2023} find that the variability in empirical emission line measurements in \MgII\, for RM160 suggest that its BLR lies farther out than the Balmer lines.

In general, we find $R_{\rm H\beta} \lesssim R_{\rm MgII } \lesssim R_{\rm H\alpha}$ for all states, as well as for the joint posterior analysis (Section~\ref{subsubsec:joint}).  The BLRs are similarly radially distributed via the shape parameter $\beta$. This regime of stratification within the BLR agrees with theoretical modeling in terms of the Balmer lines (i.e., $R_{\rm H\beta} \lesssim R_{\rm H\alpha}$), but not the placement of \MgII. There are many factors contributing to this discrepancy. The lack of variability and response in conjunction with the longer lag from \MgII\, makes reverberation analysis particularly difficult. A longer baseline of observations may be needed to properly capture the reverberating signal within the data and thus within the model. Secondly, the scenario within RM160's BLR is atypical as it has a significant contribution from non-virial orbits, meaning the structure of its BLR may be different from the models previously explored. While this conclusion disagrees with photoionization models, it agrees with prior RM results, showing that lags from \halpha\, and \MgII\, are larger than those from \hbeta, but not a distinct difference in lags between \halpha\, and \MgII\, \citep{HomayouniEtAl2020, ShenEtAl2024}. 


Future modeling efforts may need to consider more detailed radiative transfer effects to understand the relative distributions of the Balmer lines and \MgII. For instance, the emissivity-weighted lag of the BLR is dominated by the highly variable gas closest to the SMBH \citep{KoristaGoad2000, GuoEtAl2020}, thus the measured lag for the BLR can depend on its radial and azimuthal gas distribution. Other unconsidered environmental conditions may also be present. Indeed, since \MgII\, is a resonance line, its optical depth is expected to be higher than the Balmer lines \citep{SunEtAl2015}. The higher number of scatterings could increase the lag, hence the apparent radius, of the \MgII\, BLR relative to the Balmer lines \citep{KoristaEtAl2004}. \citet{OBrienEtAl1995} show changing the optical depth of the model alters the stratification, which can allow $R_{\rm H\beta} \gtrsim R_{H\alpha}$. In the optically thin case, $R_{\rm H\beta} \sim R_{H\alpha} \sim R_{\rm MgII}$. The \brains\, results show that the \MgII\, BLR may be optically thin, with a transparent midplane, low clustering, and far-side preference, validating that its size is comparable to the \halpha\, BLR. More complex photoionization modeling and dynamical modeling will be needed in the future to investigate the stratification of the \MgII\, BLR in more detail.

\subsection{Line Breathing}\label{subsec:line_breathing}

Prior theoretical analysis and time-domain observations of the BLR have found an anti-correlation between the width of a given broad emission line and its flux, a phenomenon called ``line breathing'' \citep{KoristaGoad2004}. Physically, if the continuum radiation increases, it will ionize clouds farther out in the BLR, thus increasing its emissivity-weighted radius. Consequently, the width of the emission line will decrease, as the average velocity of the emitting clouds decreases due to slower orbits at larger radii from the SMBH. Different lines have been found to have different breathing properties -- \hbeta\, shows strong breathing correlations, whereas \halpha\, and \MgII\, show less to no breathing \citep{Shen2013, DexterEtAl2019a, WangEtAl2020, GuoEtAl2020}. 


While we do find changes in BLR geometry across the low and high states of RM160, more complex analyses may need to be performed to properly assess the magnitude of this change. For example, the states we use capture year-long processes in the data, even though a change in size may occur on shorter timescales. Using shorter sections of the data would allow the model to capture a change in geometry and kinematics on these short timescales in more detail, but at the price of uncertainty in both the data and the output model parameters. Introducing a time dependency into the model parameters would allow for line breathing to be properly accounted for, although there already exists a number of degeneracies between the model parameters; adding a temporal dependency may only add to the number of degeneracies between model parameters. The maximum baseline of 10 years in the observed frame for objects in the SDSS-RM field places a limit on current time-dependent dynamical modeling analysis, but future studies with longer baselines will be able to extract additional time-dependent information.

\subsection{Comparison to Prior Work}\label{subsec:comparison}
\subsubsection{Lag $\tau$}\label{subsubsec:lags}

We compare relevant physical and statistical values recovered from the \brains\, modeling to the most recent SDSS-RM analysis from \citet[hereafter \citetalias{ShenEtAl2024}]{ShenEtAl2024} in Table~\ref{tab:pubval_comp}. Integrating the transfer function $\Psi(v, \tau)$ for each model indicates that the model-produced lags are higher for \hbeta, lower for \MgII, and similar for \halpha. For the high and full states, the difference is due to the period of observations used for lag analysis in \citetalias{ShenEtAl2024} -- they utilize continuum and spectroscopic data spanning from late 2009 to early 2020, while we extend the baseline to early 2022. The \hbeta\, line breathing in the high state observed after 2020 may have influenced the measured lag, as the change in size would be the most drastic for the lines closest to the SMBH. The low state also covers a different time period than SDSS-RM, excluding the first 3-4 years of data, which is when the BLR was shrinking in response to the decreased continuum, adding a contribution from a larger BLR. In addition, the reverberation may not have been fully captured in the low state data alone for the SDSS-RM analysis. The inclusion of this high state added high amplitude variability that made the reverberation in \MgII\, more easily detectable.

Prior dynamical modeling results find agreement between the lags from a given model and previous RM studies \citep{PancoastEtAl2014, GrierEtAl2017c, LiEtAl2018, WilliamsEtAl2018a, WilliamsEtAl2020a, BentzEtAl2021a, BentzEtAl2022} and velocity-resolved RM analysis \citep{VillafanaEtAl2022a}. However, these results use the same continuum and emission line data as the RM analysis, while we employ a discrepant dataset as compared to \citetalias{ShenEtAl2024}. \citet{LiEtAl2018} report dynamical modeling lags shorter than those found in cross-correlation function (CCF) analysis, attributing the discrepancy to the width and multimodality of the output lag distribution. The distribution of lags output from CCF and {\tt PyROA} \citep{DonnanEtAl2021} for a majority of the sample in \citetalias{ShenEtAl2024} are fairly narrow and singly peaked, excluding this as a possible explanation. A true comparison necessitates RM analysis on the same dataset used here, which is not the focus of this work. 


\begin{table*}
    \centering
    \caption{Published Value Comparisons}
\begin{tabular}{|l?c|c|c?c|c|c?c|c|c|}
\hline
 & \multicolumn{3}{c?}{$\log_{10}(M_{\rm BH} / M_{\odot})$} & \multicolumn{3}{c?}{Lag [d]} & \multicolumn{3}{c|}{$\log_{10}(f)$} \\[5pt]
\hline
 & \hbeta & \halpha & \MgII & \hbeta & \halpha & \MgII & \hbeta & \halpha & \MgII \\
\hline
Low & $7.65^{+0.04}_{-0.27}$ & $7.73^{+0.15}_{-0.15}$ & $7.74^{+0.24}_{-0.00}$ & $28.76^{+0.58}_{-41.51}$ & $128.91^{+0.41}_{-2.59}$ & $103.03^{+18.97}_{-0.73}$ & $0.06^{+0.05}_{-0.06}$ & $-0.18^{+0.48}_{-0.38}$ & $-0.27^{+0.24}_{-0.03}$ \\[5pt] 
High & $7.85^{+0.40}_{-0.15}$ & $7.72^{+0.04}_{-0.40}$ & $7.95^{+0.55}_{-0.08}$ & $56.05^{+1.43}_{-14.59}$ & $90.30^{+6.40}_{-12.97}$ & $56.65^{+15.78}_{-41.21}$ & $0.14^{+0.38}_{-0.06}$ & $0.00^{+0.29}_{-0.35}$ & $0.16^{+0.21}_{-0.12}$ \\[5pt] 
Full & $7.66^{+0.21}_{-0.05}$ & $7.89^{+0.51}_{-0.00}$ & $7.63^{+0.31}_{-0.09}$ & $43.50^{+12.11}_{-23.11}$ & $104.43^{+38.64}_{-1.44}$ & $69.66^{+30.96}_{-19.23}$ & $0.07^{+0.33}_{-0.07}$ & $0.05^{+0.31}_{-0.04}$ & $-0.21^{+0.08}_{-0.06}$ \\[5pt] 
\hline
\citetalias{ShenEtAl2024} & $8.11^{+0.02}_{-0.02}$ & $8.32^{+0.02}_{-0.02}$ & $8.43^{+0.02}_{-0.02}$ & $69.79^{+2.90}_{-2.73}$ & $122.60^{+5.35}_{-5.51}$ & $131.12^{+6.39}_{-5.86}$ & \multicolumn{3}{c|}{ $0.62^{+0.32}_{-0.32}$ } \\[5pt] 
\hline
\end{tabular}
    \label{tab:pubval_comp}
\end{table*}


Fries et al. (2024, in prep.) obtain integrated time lags for the same emission lines discussed here, using a similar dataset (see Section~\ref{sec:data}), though they omit \MgII\, due to poor fitting results. For the low state, they find $\tau_{\rm H\beta} = 52.1$ d and $\tau_{\rm H\alpha} = 90.5$ d. For the high state, they find $\tau_{\rm H\beta} = 59.5$ d and $\tau_{\rm H\alpha} = 81.9$ d. We find similar lags for \hbeta\, in the high state and \halpha\, in the low state. We also observe a change in \halpha\, lag consistent with a decrease from the low state to the high state, although our uncertainties are sufficiently large that the lags from the two states are also consistent with no change or an increase. A comparison using the same definitions of states is needed for a proper comparison of lags.


\subsubsection{The Virial Factor $f$}\label{subsubsec:virial_f}

There are a plethora of assumptions and uncertainties surrounding the virial factor $f$ used for RM and single-epoch estimates of $M_{\rm BH}$. $f$ implicitly assumes viriality, so once this assumption breaks down, it is unclear how $f$ depends on the characteristics of the BLR \citep{LinzerEtAl2022}. Furthermore, viriality assumes all clouds move only under the influence of gravity from the SMBH; if another force acts on the clouds in the same manner as gravity (e.g., radiation pressure), it would cause an overestimate for $M_{\rm BH}$ and additional uncertainties \citep{Krolik2001, Shen2013}. However, it has been shown that RM estimates for $M_{\rm BH}$ are consistent with those accounting for radiation pressure \citep{NetzerMarziani2010}, and thus with most dynamical modeling estimates.

$f$ is calibrated using a variety of methods, comparing estimates of the virial product to estimates of $M_{\rm BH}$, including those derived including the $M-\sigma_*$ relation \citep{OnkenEtAl2004a, GrahamEtAl2011, ParkEtAl2012, WooEtAl2013a, HoKim2014} and dynamical modeling results \citep{PancoastEtAl2014, GrierEtAl2017c, WilliamsEtAl2018a, ShenEtAl2024}. Each method makes assumptions about the population of AGNs studied, and each produces a slightly different estimate. Additionally, most studies use an ensemble-averaged value for $f$, although $f$ can vary object-by-object \citep{LinzerEtAl2022}, and may vary over time. It is not understood what physical conditions can produce a change in $f$, if any.

\citet{WilliamsEtAl2018a} and \citet{WangEtAl2020} have shown that the line dispersion $\sigma$ is a more meaningful measure of line width, as opposed to the FWHM, to estimate $f$. For better comparison to prior RM results, we opt to use the RMS spectrum in place of the mean spectrum. Therefore, we use $f_{\sigma, \rm rms}$ (hereafter $f$) as an estimate of the virial factor. Using the integrated lag estimate from $\Psi(v,\tau)$ and the line dispersion from the RMS line profile for a given state, we estimate $f$ for each line in each state (Table~\ref{tab:pubval_comp}). In general, we obtain values of $f \in [0.5, 1.6]$. For completeness, we calculate $f$ using the FWHM and find consistency with $\sigma$, with those derived from the FWHM having larger uncertainties.


\begin{figure*}
    \centering
    \includegraphics[width=\textwidth]{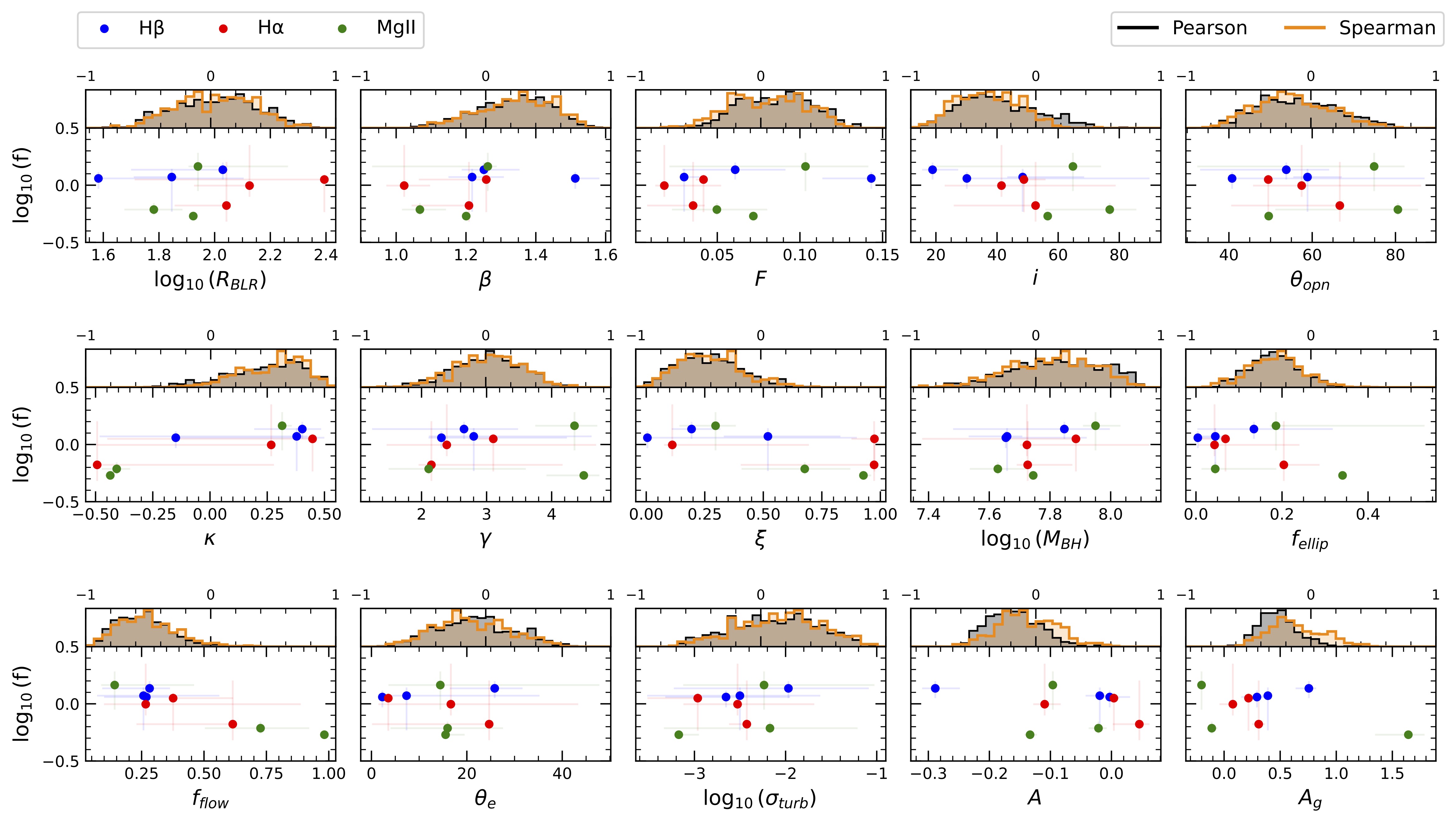}
    \caption{Correlations of the virial factor $\log_{10}(f)$ to model parameters. Values of $f$ are color-coded by their emission line. Each panel contains a scatter plot of $f$ against the model parameter (\emph{bottom}) and a histogram of the distribution for Pearson's $r$ and Spearman's $\rho$ coefficients (\emph{top}). }
    \label{fig:logf_correlations}
\end{figure*}

More than half of our values for $f$ agree with the average value from \citetalias{ShenEtAl2024}. Overall, we recover a lower value for $f$ for RM160 than the ensemble $f$ factor found in local RM AGNs, calculated by comparing mean-lag-based RM masses and dynamical-modeling masses. This discrepancy, and the span of $f$ seen across models, demonstrate that adopting the orientation-averaged $f$ factor in traditional RM can cause an overestimate of $M_{\rm BH}$ by nearly 1~dex. As the individual measurements of $f$ span a variety of inclinations, if the inclination of a given object significantly deviates from the mean inclination, so too would its virial factor \citep{McLureDunlop2002, DecarliEtAl2008, RunnoeEtAl2013, Shen2013}. Given the $1/\sin(i)$ factor between LOS velocity and true velocity, higher inclinations would require smaller virial factors to accurately measure $M_{\rm BH}$ \citep{BentzEtAl2022}. Non-virial kinematics present in the BLR, such as in RM160, further decrease $f$ due to an underestimated velocity \citep{Wandel2001}. This reasoning can be applied to the accretion state of the AGN, as well as other physical conditions such as the column density \citep{NetzerMarziani2010}. Our values are lower than the estimates using the $M-\sigma_*$ relation, including those presented for galaxies with either bulges or pseudo-bulges in \citet{HoKim2014}.

We searched for correlations between the obtained values for $f$ from each model and relevant model parameters to investigate the dependence of the virial factor on physical conditions within the BLR (Fig.~\ref{fig:logf_correlations}). We find marginal (anti-)correlations between $f$ and ($i,\theta_{\rm opn}, A, A_g$)$\beta, M_{\rm BH}$. It should be noted that $f$ is dependent on $M_{\rm BH}$, thus sharing correlations with other model parameters. Previous dynamical modeling results show an anti-correlation between $f$ and $i$ \citep{PancoastEtAl2014, WilliamsEtAl2018a, WilliamsEtAl2021}, predicted in BLR modeling \citep{GoadEtAl2012}, and a correlation between $f$ and $M_{\rm BH}$ \citep{WilliamsEtAl2018a, WilliamsEtAl2021}. Recently, \citet{VillafanaEtAl2023} investigated the correlations between model parameters within {\tt CARAMEL} and $f$, using the \citet{VillafanaEtAl2022a} sample. Similarly, they find a correlation between $f$ and $M_{\rm BH}$, an anti-correlation between $f$ and $i$, and an anti-correlation between $f$ and $\theta_{\rm opn}$. 

There are more significant correlations between $f$ and $\kappa, \xi, f_{\rm ellip}$ and $f_{\rm flow}$. These four parameters are correlated with each other, so it follows that they would all correlate with $f$. This suggests that the direction of radial cloud flow, in such a non-virial case, influences the determination of $M_{\rm BH}$. The dependence of $f$ on the state of the AGN is unknown and has varied across different dynamical modeling studies. Multiple dynamical modeling analyses find no correlation between $f$ and the Eddington ratio $\lambda_{\rm Edd}$ \citep{PancoastEtAl2014, WilliamsEtAl2018a, VillafanaEtAl2022a}, while others do \citep{GrierEtAl2017c, WilliamsEtAl2021}. \citet{LiEtAl2018} find a similar small scale factor in the super-Eddington object Mrk 142. RM160 has $\lambda_{\rm Edd} = 0.015$ \citep{ShenEtAl2019}, a relatively low accretion rate. Thus, we rule out the dependence of $f$ on the state of accretion and kinematics in RM160.

Subsequently, preferential emission toward or away from the ionizing source ($\kappa$) in conjunction with the transparency of the midplane ($\xi$) may contribute to the scale factor more so than the kinematics. A higher apparent covering factor for sources with a near-side preference will have an increased optical depth, hence an increased lag, lowering the scale factor. Most other dynamical modeling results exhibit a preference for far-side emission \citep{PancoastEtAl2014, GrierEtAl2017c, WilliamsEtAl2018a, WilliamsEtAl2020a, WilliamsEtAl2021}, or an unconstrained/varied preference \citep{LiEtAl2018, PancoastEtAl2018, BentzEtAl2021a}. Indeed, only a few objects display a near-side preference \citep{LiEtAl2018, BentzEtAl2022}.

As $f$ can vary with geometry, we investigate its change over time and between line species. \citet{WangEtAl2020} find that the relation between $M_{\rm BH}$ and luminosity $L$ changes with the observed line-breathing correlation slope $\alpha$, which differs per line. \citet{GoadEtAl2012} model a bowl-shaped BLR, and find that lines primarily located at smaller BLR radii produce larger $M_{\rm BH}$, and lower $f$.  Combining our posteriors for $f$ across states produces: $\log_{10}(f_{\rm H\beta}) = 0.09^{+0.05}_{-0.05}$, $\log_{10}(f_{\rm H\alpha}) = 0.03^{+0.10}_{-0.21}$, and $\log_{10}(f_{\rm MgII}) = -0.23^{+0.03}_{-0.03}$. $f_{\rm H\beta}$ and $f_{\rm H\alpha}$ match within their uncertainties, primarily due to similar geometries. \MgII\, displays a different geometry in terms of inclination and cloud weighting, yielding a lower $f$. Combining the posteriors across lines:  $\log_{10}(f_{\rm low}) = -0.03^{+0.03}_{-0.04}$,
$\log_{10}(f_{\rm high}) = 0.16^{+0.08}_{-0.09}$, and 
$\log_{10}(f_{\rm full}) = -0.16^{+0.23}_{-0.10}$. The full state posterior is multimodal and broad, so we caution against using this value. There is a significant increase in the scale factor from the low state to the high state due to the lower inclination and an overall higher preference for the far-side of the disk over the near-side. Indeed, Fries et al. (2024, in prep.) find that the virial product for RM160 changes over time. We find that $f$ can change across different emission lines and over time, if the geometry and/or anisotropy of the BLR change.

\subsubsection{$M_{\rm BH}$}\label{subsubsec:mbh}

Our values for $M_{\rm BH}$ are lower than those found for RM160 in \citetalias{ShenEtAl2024}. The main reason for this difference is that the average $f$ factor used in \citetalias{ShenEtAl2024} is an overestimate for RM160, incorrect with respect to RM160's geometry inferred from dynamical modeling. While the RM lag differs from the dynamical modeling lag, the difference in $f$ is large enough that it dominates the $M_{\rm BH}$ discrepancy. This result reinforces the fact that the dispersion in individual $f$ factors contributes significantly to the uncertainty in mean-lag-based RM $M_{\rm BH}$ estimates \citepalias[e.g.,][]{ShenEtAl2024}.

The BLR for RM160 contains a significant amount of non-virial motion, violating the assumptions present in traditional RM. In this case, methods such as traditional RM and velocity-resolved RM break down, producing incorrect lags, $f$ factors, and $M_{\rm BH}$. This was recently confirmed by Fries et al. (2024, in prep), who report a changing virial product for RM160. The measured lag from \citetalias{ShenEtAl2024}, and thus $M_{\rm BH}$, differ slightly due to differences in light curves used, but this difference is much smaller than the discrepancy in virial factor of nearly 1~dex. Traditional RM and other commonly used methods will provide incorrect estimates of important BLR and AGN properties in objects such as RM160. In non-virial cases, methods such as dynamical modeling where $M_{\rm BH}$ is measured independent of geometric and kinematic assumptions in the BLR will be crucial.

\section{Conclusions}\label{sec:conclusion}

We perform dynamical modeling analysis using \brains\, and the model presented in \citetalias{PancoastEtAl2014b} on a highly variable quasar in the SDSS-RM sample, RM160, using photometric and spectroscopic data spanning over a decade. We analyze the low state, high state, and combined full state of broad emission in \hbeta, \halpha, and, for the first time, \MgII. Our main findings are as follows:

\begin{itemize}
    \item The obtained mass of the SMBH in RM160 is consistent across model runs, between lines and states, granting joint posterior estimates of: $\log_{10}(M_{BH} / M_{\odot}) = $ $7.70^{+0.16}_{-0.04}$ (low state), $7.76^{+0.30}_{-0.16}$ (high state), and $7.66^{+0.12}_{-0.13}$ (full state).
    \item All models show a thick disk, inclined moderately edge-on towards the observer. The inclination of all BLRs are similar, except for \MgII, whose BLR is more edge-on than the Balmer lines. Kinematics of clouds within the BLR are mostly non-virial for all lines, with most orbits being non-elliptical, though close to Keplerian. A significant fraction of non-virial orbits are unbound, with velocity distributions centered around the escape velocity. 
    \item Owing to a degeneracy between the binary inflow/outflow parameter and certain asymmetry parameters, the direction of radial motion within the BLR as a whole is unconstrained. All \hbeta\, models are firmly inflow, although results vary between models for \halpha\, and \MgII. Discrepant kinematics across lines may suggest transient weather within the BLR, or unincluded photoionization physics within the dynamical model.
    \item Generally, $R_{\rm H\beta} \lesssim R_{\rm MgII } \lesssim R_{\rm H\alpha}$ for all states. This result is consistent with prior RM analysis, but not with BLR stratification seen in photoionization modeling. More complex photoionization physics within dynamical modeling are needed to truly resolve this discrepancy. 
    \item In all lines, $R_{\rm low} \lesssim R_{\rm high}$, indicative of line breathing. The BLR radius for \hbeta\, (significantly) and \halpha\, (marginally) vary more with time than the BLR radius for \MgII. This result is consistent with prior photoionization modeling and RM analysis of line breathing, where the Balmer lines are shown to breathe more than \MgII, if \MgII\, breathes at all.
    \item We calculate values for the virial factor $f$ for all models, and find nearly half of our models produce $f$ consistent with prior dynamical modeling results, the rest being significantly lower. There are correlations between $f$ and a number of model parameters, indicating that $f$ varies with the geometry of the BLR. Subsequently, we find that $f$ varies over time and between the different emission lines. 
    \item Our model-predicted values for $\tau$ and $M_{\rm BH}$ are lower than the estimates provided in \citetalias{ShenEtAl2024}, given the significant deviation of the specific virial factor $f$ in RM160 from the average $f$ factor. This discrepancy in $M_{\rm BH}$ may be up to $\sim 1$~dex, demonstrating that dynamical modeling is essential in recovering AGN and BLR properties in inclined, non-virial BLR scenarios.
\end{itemize}

The quality and cadence of spectroscopic data afforded by large sky surveys such as SDSS-V, combined with long-baseline programs such as SDSS-RM and BHM-RM, are able to produce in-depth physically-motivated geometric and kinematic models of the BLR for a wide range of AGN parameters. As SDSS-V continues to monitor fields with the BHM-RM program, a number of variable AGNs with high-cadence and long-baseline data will be of high enough quality to map the variability of the BLR. Increasing the sample of dynamically modeled AGNs is key to improving our understanding of its underlying physics, measuring properties of the SMBH, and verifying assumptions made in other areas of AGN analysis, especially RM and velocity-resolved RM. 

This work aids in the effort to generate a cohesive picture of the BLR across emission lines and across decades-long periods in time, and demonstrates the utility of large sky surveys such as SDSS-V in doing so. Future work within BHM-RM and other RM programs will allow construction of a sample of intensely monitored AGNs to verify and improve upon the results and interpretations seen in dynamical modeling analysis.



We thank Yan-Rong Li for help in using \brains\, and useful conversations. Z.S. and Y.S. acknowledge support from NSF grant AST-2009947. C.R. acknowledges support from Fondecyt Regular grant 1230345 and ANID BASAL project FB210003.

Funding for the Sloan Digital Sky Survey V has been provided by the Alfred P. Sloan Foundation, the Heising-Simons Foundation, the National Science Foundation, and the Participating Institutions. SDSS acknowledges support and resources from the Center for High-Performance Computing at the University of Utah. SDSS telescopes are located at Apache Point Observatory, funded by the Astrophysical Research Consortium and operated by New Mexico State University, and at Las Campanas Observatory, operated by the Carnegie Institution for Science. The SDSS web site is \url{www.sdss.org}.

SDSS is managed by the Astrophysical Research Consortium for the Participating Institutions of the SDSS Collaboration, including the Carnegie Institution for Science, Chilean National Time Allocation Committee (CNTAC) ratified researchers, Caltech, the Gotham Participation Group, Harvard University, Heidelberg University, The Flatiron Institute, The Johns Hopkins University, L'Ecole polytechnique f\'{e}d\'{e}rale de Lausanne (EPFL), Leibniz-Institut f\"{u}r Astrophysik Potsdam (AIP), Max-Planck-Institut f\"{u}r Astronomie (MPIA Heidelberg), Max-Planck-Institut f\"{u}r Extraterrestrische Physik (MPE), Nanjing University, National Astronomical Observatories of China (NAOC), New Mexico State University, The Ohio State University, Pennsylvania State University, Smithsonian Astrophysical Observatory, Space Telescope Science Institute (STScI), the Stellar Astrophysics Participation Group, Universidad Nacional Aut\'{o}noma de M\'{e}xico, University of Arizona, University of Colorado Boulder, University of Illinois at Urbana-Champaign, University of Toronto, University of Utah, University of Virginia, Yale University, and Yunnan University.

The Pan-STARRS1 Surveys (PS1) and the PS1 public science archive have been made possible through contributions by the Institute for Astronomy, the University of Hawaii, the Pan-STARRS Project Office, the Max-Planck Society and its participating institutes, the Max Planck Institute for Astronomy, Heidelberg and the Max Planck Institute for Extraterrestrial Physics, Garching, The Johns Hopkins University, Durham University, the University of Edinburgh, the Queen's University Belfast, the Harvard-Smithsonian Center for Astrophysics, the Las Cumbres Observatory Global Telescope Network Incorporated, the National Central University of Taiwan, the Space Telescope Science Institute, the National Aeronautics and Space Administration under Grant No. NNX08AR22G issued through the Planetary Science Division of the NASA Science Mission Directorate, the National Science Foundation Grant No. AST-1238877, the University of Maryland, Eotvos Lorand University (ELTE), the Los Alamos National Laboratory, and the Gordon and Betty Moore Foundation.

Based on observations obtained with the Samuel Oschin Telescope 48-inch and the 60-inch Telescope at the Palomar  Observatory as part of the Zwicky Transient Facility project. ZTF is supported by the National Science Foundation under Grants No. AST-1440341 and AST-2034437 and a collaboration including current partners Caltech, IPAC, the Oskar Klein Center at Stockholm University, the University of Maryland, University of California, Berkeley , the University of Wisconsin at Milwaukee, University of Warwick, Ruhr University, Cornell University, Northwestern University and Drexel University. Operations are
conducted by COO, IPAC, and UW.

The {\tt ztfquery} code was funded by the European Research Council (ERC) under the European Union's Horizon 2020 research and innovation programme (grant agreement n°759194 - USNAC, PI: Rigault).


%

\vspace{5mm}
\facilities{ Sloan(APO), CFHT, Bok }


\software{ {\tt astropy} \citep{astropy1, astropy2, astropy3},
\brains\, \citep{LiEtAl2018},
{\tt CDNest} \citep{Li2018},
{\tt LinMix} \citep{Kelly2007a},
{\tt matplotlib} \citep{matplotlib},
{\tt numba} \citep{numba},
{\tt numpy} \citep{numpy},
{\tt Plotly} \citep{Plotly},
{\tt PyCALI} \citep{LiEtAl2014},
{\tt scipy} \citep{scipy},
{\tt ztfquery} \citep{Rigault2018} }



\appendix

\section{Model Caveats}\label{sec:caveats}

There are a number of caveats to the dynamical modeling method, described throughout previous studies. Here, we provide a brief overview of certain caveats to note:

\begin{figure*}
    \centering
    \includegraphics[width=\textwidth]{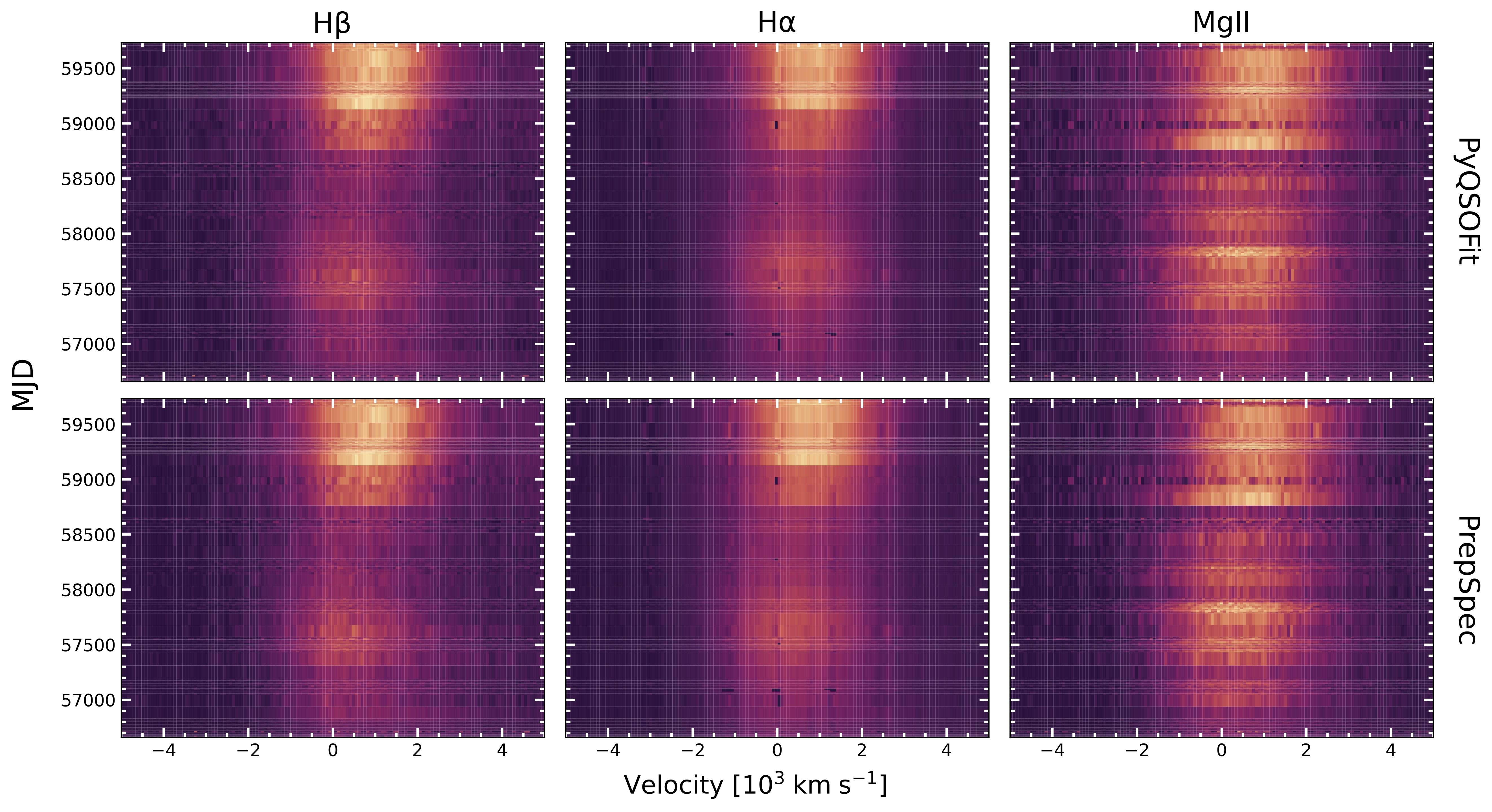}
    \caption{A comparison of the input data to \brains\, using two different spectral reduction methods for each line: \qsofit\, (top), \ps\, (bottom). \textbf{Note:} The scale for each panel is different.}
    \label{fig:input_method_comp}
\end{figure*}

\begin{itemize}
    \item The modeling performed does not model the underlying BLR gas distribution, but instead the BLR emissivity distribution. There may be non-emissive gas that contributes minimally to the observed emission line profiles, but significantly to the geometry or dynamics of the BLR. To fully model the BLR gas would require more complex radiative transfer and dynamics, similar to what has been done with {\tt CARAMEL-gas} \citep{WilliamsTreu2022}.
    \item The model geometry is axisymmetric, not accounting for azimuthal asymmetries such as ``hotspots''. It also doesn't account for effects such as self-shielding, where clouds can block emission from other clouds along the LOS. 
    \item While we adopted the model described in \citetalias{PancoastEtAl2014b}, other models may fit RM160's variability better, such as the two-zone model used in \citet{LiEtAl2018}.
    \item The radiative transfer within the model is simplistic. While the relationship between continuum and emission line variability is allowed to be nonlinear, there remain a number of assumptions made about the radiative transfer within the model. For example, all lines are treated the same, and all lines are fit independently. Careful consideration of photoionization effects needs to be taken to fully capture the physical processes within the BLR. 
    \item The model assumes temporal stability, in that the dynamics and geometry do not change over the baseline of observations/modeling. Effects of global changes in BLR structures over a long-term monitoring period, transient BLR ``weather'', such as the ``BLR holiday'' seen in NGC 5548 \citep{GoadEtAl2016}, and moving azimuthal asymmetries \citepalias{FriesEtAl2023} would manifest in unforeseen ways in the model parameters . 
    \item There are correlations among model parameters that are aphysical, in order to reproduce the input emission line shapes and variations. One in particular is the correlation between $i$ and $\theta_{\rm opn}$ seen in all previous dynamical modeling analyses. As stated in \citet{GrierEtAl2017c}, to produce singly peaked emission line profiles, the model requires $\theta_{\rm opn} \gtrsim i$, placing a prior on the opening angle and introducing a degeneracy. Additionally, there are a number of combinations of model parameters that can produce a single set of emission line shapes and variations, such as the correlation shown between $\kappa$ and $f_{\rm flow}$ and $\xi$ (see Section \ref{subsec:model_interp}).
    \item The model assumes a point-like source for the continuum photoionizing radiation, for which we use optical flux as a proxy. There have been lags found between the UV continuum and optical emission within the accretion disk in a number of AGNs \citep{CackettEtAl2007, CackettEtAl2018a, EdelsonEtAl2015a, EdelsonEtAl2017a, FausnaughEtAl2016b, SergeevEtAl2005a, McHardyEtAl2014}. The emission in the UV and optical originate from different points in space and time, with the optical emission being a smoothed, lagged version of the UV. Even the UV emission itself will not be from a point-like source, although the scale of its site of emission relative to the size of the BLR will be smaller. \citet{WilliamsEtAl2018a} perform dynamical modeling of \hbeta\, in NGC 5548, using both UV and optical continuum light curves. They report that while a marginal effect, the size of the BLR recovered depends on the continuum used, with the UV-measured BLR being larger than the optical-measured one. This selection also impacts the lag measured via dynamical modeling, with $M_{BH}$ being underestimated when using the optical light curve as a proxy. It may be that most AGNs studied have sufficiently large BLRs that the difference of lags using the UV and optical continua becomes inconsequential.
    \item Different spectral reduction choices can induce uncertainty into model parameter fits, changing the interpretation of the best-fit model. \citet{PancoastEtAl2018} performed dynamical modeling of \hbeta\, on Arp 151 excluding different regions of the broad emission line (i.e., the red wing, blue wing, and core). Excluding a given part of the emission line can increase the uncertainty in a given parameter by 50\% if not more. While the best-fit parameters between these different models are formally consistent, the lack of constraint due to large uncertainties can impact the model interpretation. Therefore, the extent of the line analyzed and the contamination from other parts of the spectrum (e.g., \FeII\, and host-galaxy emission) can affect dynamical modeling results.
\end{itemize}

\section{Comparison of Spectral Reduction Methods}\label{sec:method_comp}

\begin{figure*}
    \centering
    \includegraphics[width=\textwidth]{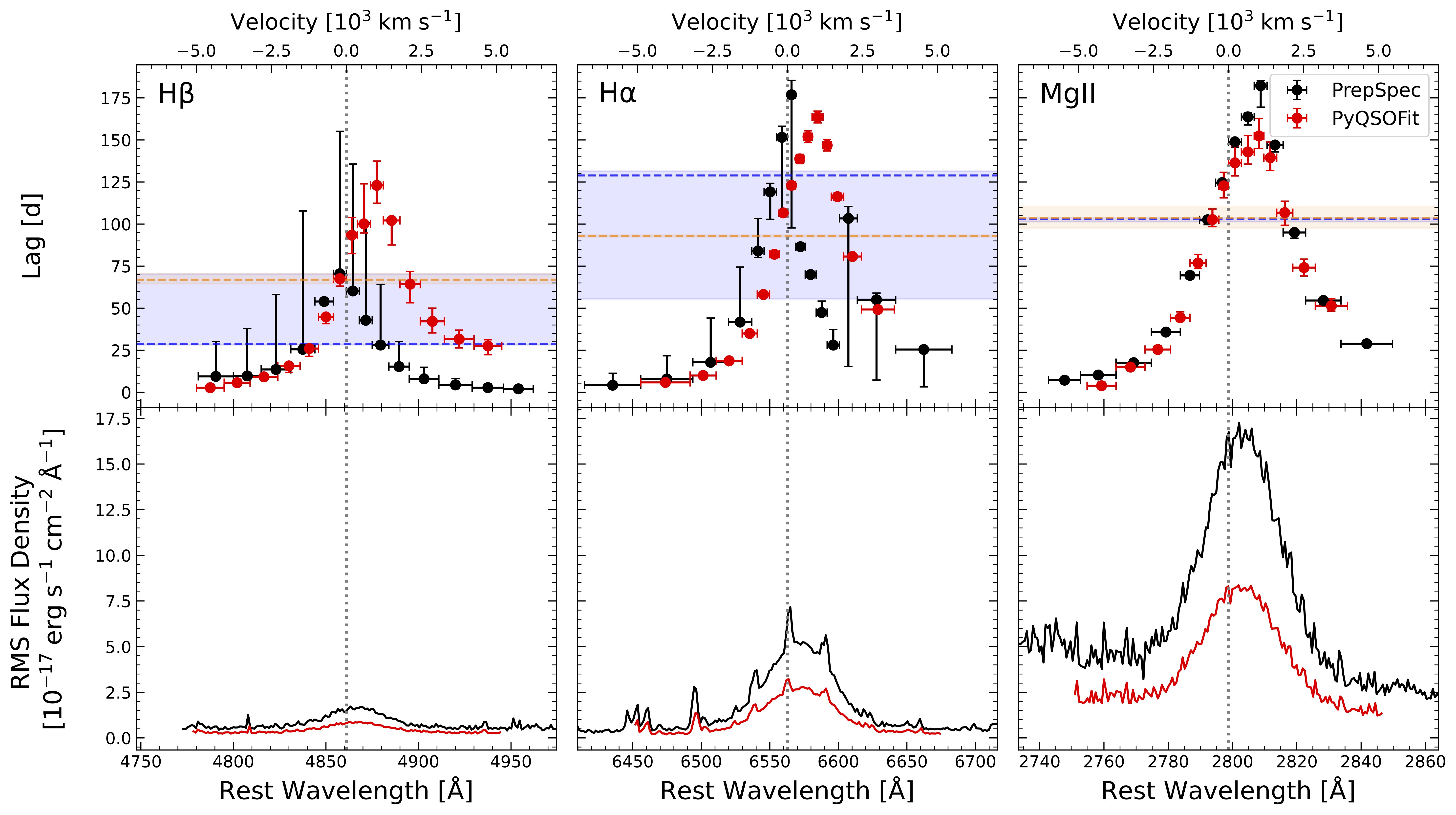}
    \caption{\emph{Top:} A comparison of the velocity-resolved lags from the \brains\, runs using the \qsofit-reduced (black) and \ps-reduced (red) spectra for each line in the low state. The blue (orange) horizontal lines and shaded regions represent the integrated lag and its uncertainties for the \qsofit\; (\ps) data. \emph{Bottom:} The RMS spectrum for each line in the low state using the \qsofit-reduced and \ps-reduced spectra.}
    \label{fig:lagcomp_low}
\end{figure*}

We now compare the dynamical modeling results for the \ps-reduced and \qsofit-reduced spectra. The spectra are processed with \qsofit\, using the procedure described in Section~\ref{subsec:spectroscopy}, and with \ps\, using standard settings, iterating until $\chi^2_\nu \sim 1$. As a check, we compare the output light curves and emission line profiles for each of the two software packages for each line and state, and find good agreement. The main difference between the two output line profiles result from the continuum subtraction, with \ps\, subtracting more than \qsofit. \citetalias{FriesEtAl2023} note that using \qsofit\, to extract line profiles produces a larger line width, lower line shift, and larger \hbeta\, flux than their method of Gaussian fitting. We find that \ps\, produces lower flux for all lines.

We perform dynamical modeling with two sets of spectra for all lines for both the low state and the high state. The sets of parameters output from the \ps-reduced spectra are generally similar to those for the \qsofit-spectra, within their uncertainties. Owing to the degeneracy in parameters, and the fact that multiple combinations of model parameters can produce a given continuum and emission line profile, some best-fit \qsofit\, models are different kinematically from their \ps\, counterparts. This behavior is similar to the result found in \citet{PancoastEtAl2018}, who state that the portion of the line studied, as well as the reduction used to obtain a given line profile, can increase the uncertainties in model parameters (see Appendix \ref{sec:caveats}).

\begin{figure}
    \centering
    \includegraphics[width=\textwidth]{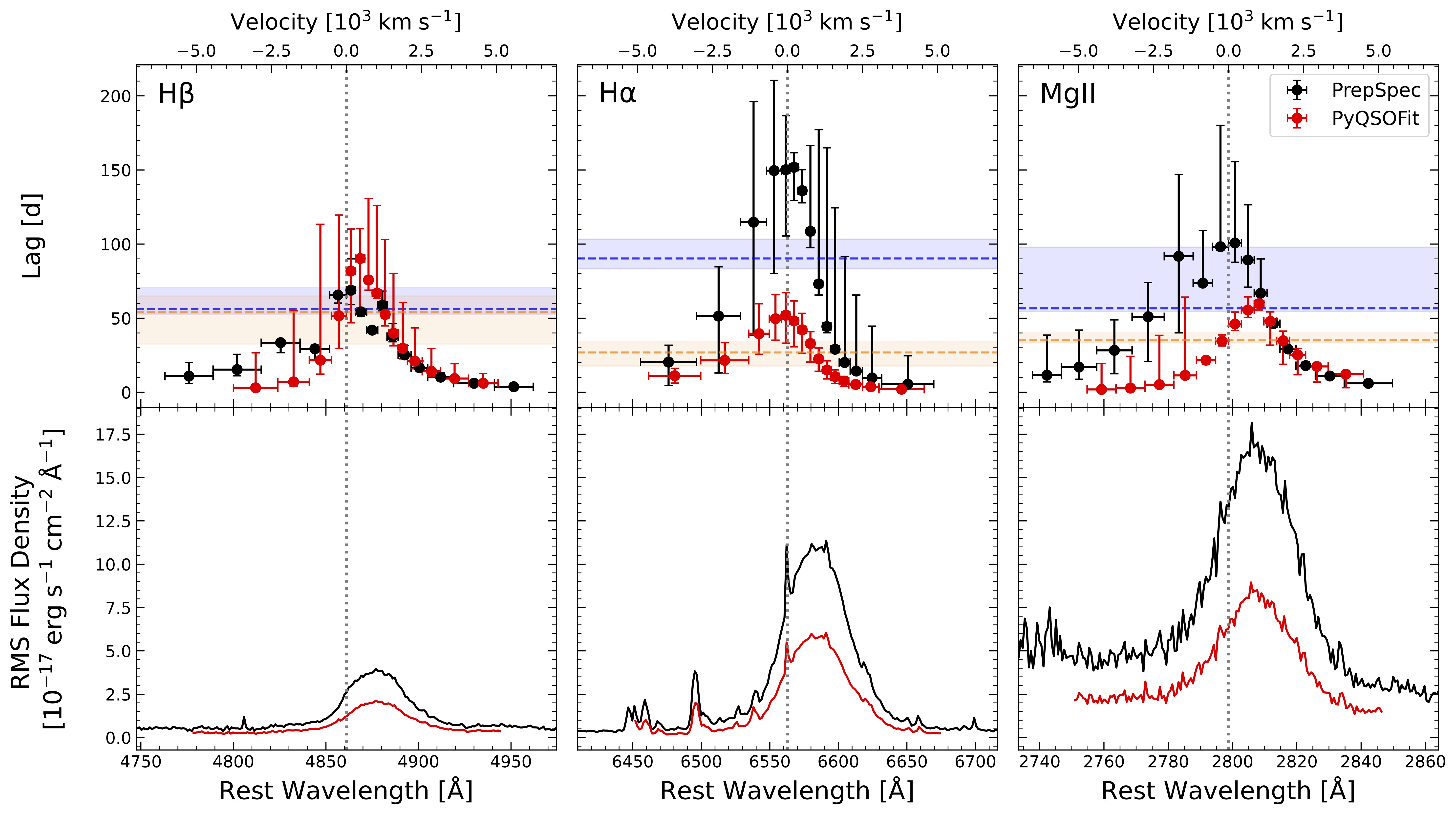}
    \caption{Same as Fig.~\ref{fig:lagcomp_low}, but for the high state.}
    \label{fig:lagcomp_high}
\end{figure}

To demonstrate this effect, we compare the output velocity-resolved lag profiles from each line for each state, comparing the \qsofit\, and \ps\, results in Figs.~\ref{fig:lagcomp_low} \& \ref{fig:lagcomp_high}. The patterns are similar between the two models for most of the lines/states. The profiles for \hbeta\, and \halpha\, with \ps\, in the low state agree with \qsofit\, in the blue wing, but peak in the red wing instead of the blue wing, indicative of outflow instead of inflow. The \MgII\, profiles between the two software packages in the low state match almost identically, suggesting outflow. The integrated lags between \qsofit\, and \ps\, are in agreement for all lines in the low state. The high state results for \ps\, deviate more from the \qsofit\, results, showing shorter lags overall, although the shapes of the lag profiles match better. In particular, \hbeta\, and \halpha\, both peak in the red wing for both packages. \MgII\, differs the most, showing outflow for \ps\, and inflow for \qsofit. Additionally, the integrated lags for \halpha\, and \MgII\, do not match within their uncertainties. This result may be due to the increased continuum subtraction dampening the variability in the emission line profile, although the calibration factor used for both methods is the same. Similar comparisons are seen when inspecting the model parameters. Thus, \ps\, produces results similar to \qsofit\, within uncertainty, accounting for model degeneracies and caveats.

\section{Comparison of Virial Factors}\label{sec:method_comp}


\begin{table*}
    \centering
    \caption{Correlations with the Virial Factor}
    \begin{adjustwidth}{-1cm}{}
\begin{tabular}{|l?c|c|c?c|c|c|}
\hline
 & \multicolumn{3}{c?}{$\log_{10}(f_{\sigma, \rm rms})$} & \multicolumn{3}{c|}{$\log_{10}(f_{\rm FWHM, rms})$} \\[5pt]
\hline
\textbf{Parameter} & \textbf{Slope} & \textbf{Pearson $r$} & \textbf{Spearman $\rho$} & \textbf{Slope} & \textbf{Pearson $r$} & \textbf{Spearman $\rho$} \\[5pt]
\hline
$\log_{10}(R_{BLR})$ & $0.65^{+4.68}_{-3.03}$ & $0.01^{+0.34}_{-0.37}$ & $0.02^{+0.28}_{-0.35}$ & $0.45^{+4.20}_{-4.33}$ & $-0.18^{+0.29}_{-0.28}$ & $-0.13^{+0.28}_{-0.25}$ \\[5pt] 
$\beta$ & $0.52^{+0.67}_{-0.69}$ & $0.24^{+0.30}_{-0.27}$ & $0.23^{+0.28}_{-0.28}$ & $1.45^{+0.88}_{-0.90}$ & $0.45^{+0.25}_{-0.22}$ & $0.42^{+0.29}_{-0.24}$ \\[5pt] 
$F$ & $0.38^{+2.87}_{-3.10}$ & $0.15^{+0.28}_{-0.27}$ & $0.08^{+0.32}_{-0.28}$ & $2.99^{+3.91}_{-4.18}$ & $0.33^{+0.25}_{-0.24}$ & $0.22^{+0.30}_{-0.23}$ \\[5pt] 
$i$ & $-0.01^{+0.01}_{-0.01}$ & $-0.30^{+0.30}_{-0.31}$ & $-0.37^{+0.28}_{-0.27}$ & $-0.01^{+0.01}_{-0.01}$ & $-0.44^{+0.22}_{-0.32}$ & $-0.50^{+0.20}_{-0.27}$ \\[5pt] 
$\theta_{opn}$ & $0.00^{+0.16}_{-0.26}$ & $-0.12^{+0.29}_{-0.36}$ & $-0.12^{+0.32}_{-0.28}$ & $0.01^{+0.17}_{-0.15}$ & $-0.24^{+0.25}_{-0.34}$ & $-0.24^{+0.26}_{-0.29}$ \\[5pt] 
$\kappa$ & $0.48^{+0.16}_{-0.18}$ & $0.47^{+0.41}_{-0.28}$ & $0.50^{+0.33}_{-0.27}$ & $0.55^{+0.28}_{-0.28}$ & $0.29^{+0.28}_{-0.22}$ & $0.33^{+0.27}_{-0.27}$ \\[5pt] 
$\gamma$ & $-0.08^{+0.19}_{-0.20}$ & $0.02^{+0.25}_{-0.26}$ & $0.05^{+0.28}_{-0.23}$ & $-0.11^{+0.26}_{-0.26}$ & $-0.13^{+0.22}_{-0.29}$ & $-0.08^{+0.28}_{-0.27}$ \\[5pt] 
$\xi$ & $-0.42^{+0.26}_{-0.23}$ & $-0.42^{+0.25}_{-0.29}$ & $-0.40^{+0.27}_{-0.27}$ & $-0.55^{+0.34}_{-0.32}$ & $-0.50^{+0.20}_{-0.25}$ & $-0.50^{+0.20}_{-0.25}$ \\[5pt] 
$\log_{10}(M_{BH})$ & $1.16^{+2.34}_{-1.62}$ & $0.25^{+0.45}_{-0.33}$ & $0.15^{+0.32}_{-0.32}$ & $1.06^{+2.69}_{-2.20}$ & $0.08^{+0.30}_{-0.30}$ & $0.05^{+0.33}_{-0.25}$ \\[5pt] 
\hline
$f_{ellip}$ & $-0.93^{+0.78}_{-0.66}$ & $-0.28^{+0.23}_{-0.20}$ & $-0.27^{+0.23}_{-0.23}$ & $-1.22^{+1.13}_{-1.06}$ & $-0.30^{+0.19}_{-0.20}$ & $-0.27^{+0.22}_{-0.22}$ \\[5pt] 
$f_{flow}$ & $-0.52^{+0.20}_{-0.20}$ & $-0.59^{+0.16}_{-0.23}$ & $-0.57^{+0.20}_{-0.28}$ & $-0.72^{+0.36}_{-0.34}$ & $-0.54^{+0.19}_{-0.25}$ & $-0.50^{+0.23}_{-0.28}$ \\[5pt] 
$\theta_{e}$ & $-0.01^{+0.02}_{-0.02}$ & $-0.09^{+0.30}_{-0.30}$ & $-0.11^{+0.28}_{-0.28}$ & $-0.01^{+0.02}_{-0.02}$ & $-0.14^{+0.26}_{-0.32}$ & $-0.15^{+0.28}_{-0.27}$ \\[5pt] 
$\log_{10}(\sigma_{turb})$ & $0.25^{+0.32}_{-0.37}$ & $0.13^{+0.38}_{-0.36}$ & $0.12^{+0.32}_{-0.35}$ & $0.36^{+0.45}_{-0.50}$ & $0.10^{+0.39}_{-0.36}$ & $0.13^{+0.37}_{-0.35}$ \\[5pt] 
\hline
$A$ & $-0.50^{+1.00}_{-1.02}$ & $-0.25^{+0.21}_{-0.24}$ & $-0.13^{+0.22}_{-0.25}$ & $-0.53^{+1.33}_{-1.22}$ & $-0.08^{+0.19}_{-0.20}$ & $0.03^{+0.23}_{-0.25}$ \\[5pt] 
$A_g$ & $-0.11^{+0.17}_{-0.17}$ & $-0.26^{+0.15}_{-0.14}$ & $-0.13^{+0.20}_{-0.22}$ & $-0.08^{+0.25}_{-0.25}$ & $-0.18^{+0.17}_{-0.18}$ & $0.07^{+0.23}_{-0.18}$ \\[5pt]  
\hline
\end{tabular}
\end{adjustwidth}
    \tablecomments{The slope shown represents the slope obtained when assuming a linear relationship between $\log_{10}(f)$ and the model parameter, using the LinMix algorithm \citep{Kelly2007a}. The correlation coefficients are the same as those shown in Fig.~\ref{fig:logf_correlations}.}
    \label{tab:fcorr}
\end{table*}


\begin{figure}
    \centering
    \includegraphics[width=0.5\linewidth]{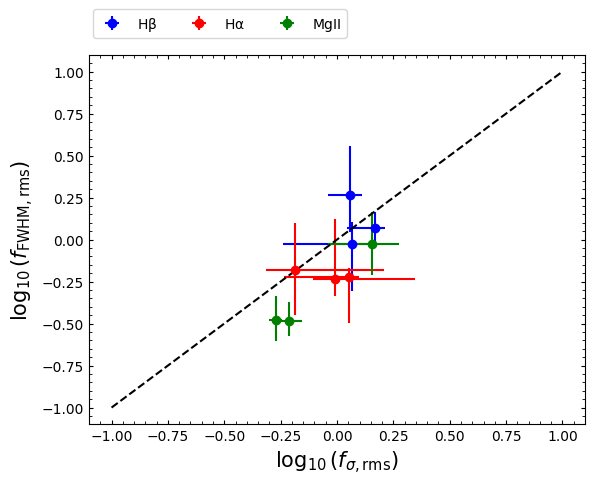}
    \caption{Comparison of $f_{\rm FWHM, rms}$ and $f_{\sigma, \rm rms}$. }
    \label{fig:enter-label}
\end{figure}


\bibliography{Refs}{}
\bibliographystyle{aasjournal}



\end{document}